\documentclass[prd,onecolumn,nopacs,nofootinbib,preprintnumbers]{revtex4}

\usepackage{amsmath}
\usepackage{graphicx}
\usepackage{amsfonts}
\usepackage{latexsym}
\usepackage{bbold}
\usepackage{wasysym}
\usepackage{comment}
\usepackage{ tipa } 
\usepackage[polish,greek,spanish,english]{babel}

\newcommand{\be}{\begin{equation}}
\newcommand{\ee}{\end{equation}}
\newcommand{\beq} {\begin{equation}}
\newcommand{\eeq} {\end{equation}}
\newcommand{\ba}{\begin{eqnarray}}
\newcommand{\ea}{\end{eqnarray}}
\newcommand{\lp}{\left(}
\newcommand{\rp}{\right)}
\newcommand{\lb}{\left[}
\newcommand{\rb}{\right]}

\newcommand{\e}{\mathrm{e}}
\newcommand{\ie}{\text{\textschwa}}

\newcommand{\diff}{{{\rm d}}}

\begin{document}

\title{Scale transformations in metric-affine geometry}

\author{Damianos Iosifidis}
\affiliation{Institute of Theoretical Physics, Department of Physics
Aristotle University of Thessaloniki, 54124 Thessaloniki, Greece}
\email{diosifid@auth.gr}
\author{Tomi Koivisto}
\affiliation{Nordita, KTH Royal Institute of Technology and Stockholm University, Roslagstullsbacken 23, SE-10691 Stockholm, Sweden}
\affiliation{National Institute of Chemical Physics and Biophysics, R\"avala pst. 10, 10143 Tallinn, Estonia}
\email{tomi.koivisto@nordita.org}

\preprint{NORDITA-2018-109}

\date{\today}
\begin{abstract}

This article presents an exhaustive classification of metric-affine theories according to their scale symmetries. 
First it is clarified that there are three relevant definitions of a scale transformation. These correspond to a projective transformation of the connection, a rescaling of the orthonormal frame,
and a combination of the two. The most general second order quadratic metric-affine action, including the parity-violating terms, is constructed in each of the three cases. 
The results can be straighforwardly generalised by including higher derivatives, and implemented in the general metric-affine, teleparallel, and symmetric teleparallel geometries.

\end{abstract}

\maketitle

\allowdisplaybreaks


\tableofcontents

\section{Introduction}
\label{intro}


The concept of gauge transformation was introduced into physics a century ago in the work of Hermann Weyl \cite{Weyl:1918ib} which generalised the spacetime 
geometry in Einstein's recent General Theory of Relativity (GR) by incorporating the relativity of magnitudes. The (pseudo)-Riemannian geometry of GR describes a
curved spacetime, wherein the direction of a vector parallel transported around a closed loop can rotate. In Weyl geometry, the magnitude of the vector can also change.
The theory was made to be invariant under the simultaneous transformation of the gravitational metric and the electromagnetic field, which was the basis for a geometrical
unification of the two fundamental interactions known at the time.  

About a decade later it was clarified that the electromagnetic interaction requires rather the compact gauge group $U(1)$ \cite{Weyl:1929fm}, and the extension of the gauge
principle to non-Abelian groups has eventually lead to the great successes of the modern standard model of particle physics some that begun about a fifty years ago 
\cite{ORaifeartaigh:1997dvq}. Thus, the very foundation of current theoretical physics can be traced back to the geometrical analysis of GR, and at its the core lie the issues of scales and integrability \cite{Koivisto:2018aip}. As expressed in Ref. \cite{Blagojevic:2013xpa}: ``It was Weyl's desire to remove all elements of an action at a distance theory from geometry. The direction of a vector in Riemannian geometry became nonintegrable, but its length remains integrable. This situation Weyl wanted to change. Weyl's interpretation of the Weyl covector $Q$ as electromagnetic potential turned out not to be viable - basically because the electric charge has no intrinsic relation to the geometry of spacetime - but the geometry Weyl created will reappear as linked to the gauge theory of scale transformations''. 
Currently, there is still no consensus on the status of scale invariance in physics, but a natural viewpoint is to regard it as a fundamental, exact symmetry which is broken in a 
spontaneous fashion \cite{THooft:2015jcw}. For reviews on various aspects of scale invariance, see \cite{Hehl:1994ue,Mannheim:2011ds,Scholz:2011za,Nakayama:2013is,Rachwal:2018gwu}.

The aim of this paper is to study scale invariance and closely related symmetries in the general geometrical framework of metric-affine gravity \cite{Hehl:1994ue}.  We work mostly in the so called Palatini formalism, where the independent gravitational variables are the metric and the affine connection \cite{Olmo:2011uz,Heisenberg:2018vsk}. As well-known, this formalism extends the (pseudo-)Riemannian
framework by taking into account two fundamental properties of spacetime geometry besides the (metric) curvature: torsion \cite{Shapiro:2001rz,Hehl:2007bn} and non-metricity \cite{Vitagliano:2013rna,Koivisto:2018aip}, the former dating back to the work Cartan, the prototype manifestation of the latter being the Weyl covector $Q$. Whilst in the first formulation of GR the metric was the only variable and gravitation was attributed to its curvature, Einstein himself a few years later considered an alternative formulation in terms of torsion, and in general became to uphold the view that the spacetime connection, rather than the metric, should be regarded as the ``directly relevant conceptual element'' \cite{Blagojevic:2013xpa}. In a spacetime equipped with an independent connection, new possibilities arise to realise the transformation of physical scales. As Weyl had clarified, the connection may be retained invariant whilst the metric is transformed \cite{PhysRev.77.699}. On the other hand, the projective transformation a.k.a. the $\lambda$-transformation, 
which is a change of the connection that leaves its autoparallels invariant, was a recurrent theme in the early history of geometrical unification \cite{Goenner:2014mka}, and it indeed at the same time it is related to the gauge transformation of the electromagnetic field and the geometric construction of the famous Ehlers-Pirani-Schild method, which is basically the separation of the conformal and the projective structures on a manifold \cite{Ehlers2012}.

A specific task we are about to undertake is {the systematic classification of quadratic metric-affine theories according to their rescaling and projective symmetries}. Currently, there is a resurgence of activity
in alternative formulations of gravity theory in the context generalised geometry, especially motivated by the mysteries of modern cosmology \cite{Heisenberg:2018vsk}. 
The most extensively studied example is the $f(R)$ gravity, wherein the important role of the conformal relation has been eludicated in the metric models \cite{Magnano:1993bd,Faraoni:1998qx}, the Palatini models \cite{Allemandi:2004yx,Iosifidis:2018zjj} and their unifications \cite{Amendola:2010bk,Capozziello:2015lza}.
In a more generic context, the interest of many investigators is captured by ``the trinity of gravity'' \cite{Heisenberg:2018vsk} i.e. the facts that GR can be equivalently formulated in terms of either curvature, torsion or non-metricity, and that furthermore, each these three formulations may give rise to inequivalent generalisations, thus opening new avenues to address the cosmological problems and the more long-standing foundational issues in the theory of gravity. In this paper we also pave the way for explorations into the general teleparallel geometry that is flat but exhibits both non-metricity and torsion. Recently, interesting insights have been found into a possible
duality between the latter two, mediated via the projective transformation \cite{Jimenez:2016opp, Iosifidis:2018diy,Iosifidis:2018zjj}. The many possible 
applications of scale-invariant theories include unification \cite{Weyl:1918ib,Koivisto:2018aip}, the hierarchy problem \cite{Shaposhnikov:2008xi,Kannike:2015apa}, the cosmological constant 
problem \cite{DeMartini:2017izr,Lucat:2018slu}, dark energy \cite{Bamba:2013jqa,Momeni:2014taa,Silva:2015dea}, dark matter \cite{Mannheim:2011ds,Israelit:2002iv}, leptogenesis \cite{Lewandowski:2017wov}, wormholes \cite{Lobo:2008zu,Hohmann:2018shl}, non-singular
black holes \cite{Modesto:2016max,Bambi:2016wdn} and their evaporation \cite{Bambi:2016yne,Bambi:2017ott}, the origin of time \cite{Hazboun:2013lra}, inflation \cite{Bezrukov:2012hx,Ferreira:2016wem} and its bouncing \cite{Gegenberg:2016vhx} and other \cite{Bars:2013yba} alternatives. Previously various versions of scale invariance in metric-affine geometry have been considered in e.g. Refs. \cite{Obukhov:1982zn,Hehl:1994ue,Zhytnikov:1993ew,Shtanov:1994sh,Karahan:2012yz,Moon:2009zq,Vazirian:2013baa,Wright:2016ayu,Lucat:2016eze,Barnaveli:2018dxo}, and some recent discussions of projective invariance in metric-affine geometry are found in Refs. \cite{Jimenez:2016opp, Iosifidis:2018diy,Iosifidis:2018zjj,Afonso:2017bxr,Aoki:2018lwx,Janssen:2018exh}. 

Weyl's original theory \cite{Weyl:1918ib} was written in terms of the unique scale-invariant quadratic Riemannian curvature scalar constructed from the metric and the metrical connection. It is well-known that this term, the ``Weyl tensor squared'', is higher order in derivatives, and in fact accommodates a ghost. Dirac had 
already discarded this particular formulation of the action principle on aesthetic grounds, though he recognised Weyl's geometric unification as ``the outstanding one, unrivalled by its simplicity and beauty'' \cite{Dirac:1973gk}. Dirac nevertheless continued to work within the framework of (extended) metrical Riemannian geometry. His resolution was to introduce a compensating field, $\psi$, which has the transformation law $\psi \rightarrow e^{-\phi}\psi$ under the rescaling $g_{\mu\nu} \rightarrow e^{2\phi}g_{\mu\nu}$ of the metric\footnote{As is a usual practice, we refer to tensors by their components. Thus, e.g. the Weyl (co)vector $Q=Q_\mu \diff x^\mu$ will be referred to as $Q_\mu$ from now on.}
 $g_{\mu\nu}$, and couple this field non-minimally to the Einstein-Hilbert term. With a particular coefficient of the kinetic term for the field $\psi$, the non-minimally coupled scalar field action becomes locally scale-invariant. Nowadays the $\psi$ is often known as the dilaton, and the action of the Weyl-Dirac theory (see \cite{israelit} for an introduction) is referred to as the conformally coupled Brans-Dicke scalar-tensor theory. Here we shall follow the Dirac's approach and generalise the theory into metric-affine geometry. The task is to deduce the ``co-covariant'' terms that can be constructed from torsion and non-metricity. The first step is to catalogue all the diffeomorphism-invariant terms and the step that will follow is the construction of their scale-covariant combinations. 

We have made an attempt at a systematic and pedagogic presentation of this paper. It is therefore organised such that many detailed derivations are included in the three Appendices, which
also contain some simple illustrative examples. There is often some confusion about the definitions and especially the physical interpretations of the scale transformations, an example
being the errors that regularly arise about the (non)equivalence of the Jordan and the Einstein frames in the $f(R)$ and other scalar-tensor theories \cite{Faraoni:2006fx}, despite that
the issue was explained in exemplary clarity already in the seminal paper of Brans and Dicke \cite{Brans:1961sx}. For these reasons, we discuss the basics of a scale-invariant theory
at some length.  First, in Section \ref{framework} we elaborate on the meaningful definitions of a scale transformation, from both the perspectives of spacetime geometry (greek indices in our notation) and tangent space geometry (latin indices), and then, in Section \ref{simple} we introduce a very simple example of scale invariant theory, the ``Einstein-Cartan-Weyl-Dirac'' action,
in order to illustrate the formulation of such a theory and its physical interpretation. The main derivations and results of this paper are given Sections \ref{quadratic} and \ref{general}, wherein we deduce the generic quadratic scale-covariant theories in terms of torsion and non-metricity in the parity-even and parity-odd cases, respectively. The results will be summarised in Table \ref{theorytable}. Finally, we conclude in Section \ref{conclu} by gathering the bunches of scalars in Table \ref{table2}. 

\section{The Geometrical Framework}
\label{framework}

In this section we will define the scale transformations to be considered in the following. We consider two formulations, which are referred to as the Palatini formalism and the metric-affine gauge formalism, for reasons explained in Ref. \cite{BeltranJimenez:2018vdo}.  

\subsection{Scale Transformations in the Palatini Formalism}

In GR, the metric $g_{\mu\nu}$ is the only independent variable. The rescaling of the metric, by a coordinate-dependent function $\phi(x)$, is defined simply as
\be \label{ct}
g_{\mu\nu} \rightarrow e^{2\phi} g_{\mu\nu}\,.
\ee
The metric connection is given by the Christoffel symbol, the unique symmetric metric-compatible connection    
\be \label{christoffel}
\left\{^{\phantom{i} \alpha}_{\beta\gamma}\right\} = \frac{1}{2}g^{\alpha\lambda}\lp g_{\beta\lambda,\gamma}
+  g_{\lambda\gamma,\beta} - g_{\beta\gamma,\lambda}\rp\,.
\ee
Under the transformation (\ref{ct}), this obviously transforms as
\be \label{weyl}
\left\{^{\phantom{i} \alpha}_{\beta\gamma}\right\}  \rightarrow \left\{^{\phantom{i} \alpha}_{\beta\gamma}\right\}  +  \phi^{,\alpha}g_{\beta\lambda} - 2\delta^\alpha_{(\beta}\phi_{,\lambda)}\,.
\ee
This was the original form of the scaling transformation in Weyl's 1918 theory \cite{Weyl:1918ib}, where it was to be performed in concert with the gauge transformation of the electromagnetic field. 
One may consider the three new linear terms in the connection (\ref{weyl}) to appear with independent (constant) coefficients $w_1$, $w_2$, $w_3$,
\be \label{distortion}
\left\{^{\phantom{i} \alpha}_{\beta\gamma}\right\}  \rightarrow \left\{^{\phantom{i} \alpha}_{\beta\gamma}\right\}  +  w_1\phi^{,\alpha}g_{\beta\gamma} + w_2\delta^\alpha_{\beta}\phi_{,\gamma}
+  w_3\delta^\alpha_{\gamma}\phi_{,\beta}\,.
\ee
This generalisation of Weyl geometry can be called the linear vector distortion (where in the above, the vector field is reduced to the gradient of the scalar $\phi$), and it has been recently used to parameterise deviations from Riemannian geometry \cite{Jimenez:2014rna,Jimenez:2015fva,Jimenez:2016opp}, see also \cite{0264-9381-8-9-004,Karahan:2012yz} and \cite{Haghani:2014zra,Shahidi:2018iwm}.  

\begin{table}[h]
\begin{tabular}{ |c|c|c|c|c|} 
 \hline
transformation & $w_1$ & $w_2$ & $w_3$ & geometry \\
 \hline
 \hline
 conformal metric & $-\frac{1}{2}$ & $\frac{1}{2}$ & $\frac{1}{2}$ & Riemannian \\
 \hline
conformal MAG & $0$ & $0$ & $0$ & orthonormal\\
 \hline
projective & 0 & $1$ & 0 & holonomic \\
\hline
projective$_{//}$ & 0 & $\frac{1}{2}$ & $-\frac{1}{2}$ & teleparallel \\
\hline
projective$_{\parallel}$ & 0 & $\frac{1}{2}$ & $\frac{1}{2}$ & symmetric telep. \\
\hline 
\end{tabular}
 \caption{The projections of the affine connection  (\ref{distortion}), or (\ref{weyl2}), in terms of the vector distortion parameters in generalised Weyl geometry \cite{Jimenez:2015fva,Jimenez:2016opp}.} \label{table1}
\end{table}

In the paper at hand however, rather than employing a phenomenological parameterisation of the connection, we work in the Palatini formalism, wherein in addition to the metric, the affine
connection $\Gamma^\alpha{}_{\mu\nu}$ is an independent dynamical variable \cite{Olmo:2011uz}. The action for the gravitational theory is written in terms of both $g_{\mu\nu}$ and $\Gamma^\alpha{}_{\mu\nu}$,
and varied with respect to both of them independently. In this sense, the connection is not fixed a priori, but assumes its form dynamically, as a solution to its equation of motion. 
In such a theory, one reasonable realisation of the rescaling is simply $w_1=w_2=w_3=w_4=0$  \cite{PhysRev.77.699}, wherein
\be \label{weyl2}
\Gamma^\alpha{}_{\beta\lambda} \rightarrow \Gamma^\alpha{}_{\beta\lambda}  + w_1\phi^{,\alpha}g_{\beta\gamma} + w_2\delta^\alpha_{\beta}\phi_{,\gamma}
+  w_3\delta^\alpha_{\gamma}\phi_{,\beta} + w_4\epsilon^{\alpha\delta}{}_{\beta\gamma}\phi_{,\delta}
\ee
would be the most general linear additive transformation. The different cases considered in this paper are summarised in Table \ref{table1}.

Generically, the independent connection may have the metric-compatible piece (\ref{christoffel}), but also non-metricity and torsion which do not necessarily reduce to pure vector or scalar components. 
The conventions we shall use are such that for the covariant derivative $\nabla_\mu$ of a vector $u^\mu$ with respect to the independent connection $\Gamma^\alpha{}_{\mu\nu}$, we have 
\beq
\nabla_{\mu}u^{\lambda}=\partial_{\mu}u^{\lambda}+\Gamma^{\lambda}{}_{\nu\mu}u^{\nu}\,.
\eeq
The curvature of the connection is defined as
\be \label{riemann}
{R}^\alpha_{\phantom{\alpha}\beta\mu\nu} = 
2\partial_{[\mu} \Gamma^\alpha_{\phantom{\alpha}\lvert\beta\rvert\nu]}
+ 2\Gamma^\alpha_{\phantom{\alpha}\lambda[\mu}\Gamma^\lambda_{\phantom{\lambda}\lvert\beta\rvert\nu]}
\ee
the torsion $S_{\mu\nu}{}^{\lambda}$ of the connection is given as its antisymmetric part,
\beq
S_{\mu\nu}{}^{\lambda}=2 \Gamma^{\lambda}{}_{[\mu\nu]} \label{torsion}
\eeq
and the non-metricity $Q_{\alpha\mu\nu}$ as the covariant derivative of the metric,
\beq
Q_{\alpha\mu\nu}=-\nabla_{\alpha}g_{\mu\nu}\,. \label{q}
\eeq
One should be aware that various other conventions are used in the literature\footnote{For example, in the conventions of \cite{BeltranJimenez:2017vop,BeltranJimenez:2018vdo}, the four previous equations
would read $\nabla_{\mu}u^{\lambda}=\partial_{\mu}u^{\lambda}+\Gamma^{\lambda}{}_{\mu\nu}u^{\nu}$, 
${R}^\alpha_{\phantom{\alpha}\beta\mu\nu} = 
2\partial_{[\mu} \Gamma^\alpha_{\phantom{\alpha}\nu]\beta}
+ 2\Gamma^\alpha_{\phantom{\alpha}[\mu\lvert\lambda\rvert}\Gamma^\lambda_{\phantom{\lambda}\nu]\beta}$,
$T^{\lambda}{}_{\mu\nu}=2 \Gamma^{\lambda}{}_{[\mu\nu]}$ and $Q_{\alpha\mu\nu}=\nabla_{\alpha}g_{\mu\nu}$, respectively.}. Clearly, in $n$ dimensions the torsion tensor can have $n^2(n-1)/2$ and the non-metricity tensor $n(n+1)/2$ independent components, which together comprise the $n^3$ independent components of a fully generic affine connection. 

The geometric concepts such of angles, distances, areas and volumes are given by the metric. The special property of the Weyl rescaling (\ref{weyl2}) is that it leaves angles invariant, whereas it changes the quantities involving magnitudes, i.e. lengths and its higher-dimensional generalisations. A priori, the geometric concept of parallel transport, of moving from one spacetime point to another, is completely independent of the metric.
This is the compelling reason to adopt Palatini formalism, wherein the parallel transport is determined by the independent field $\Gamma^\alpha{}_{\mu\nu}$. In particular, the geodesic equation that defines the parallel transport of a vector $u^\mu$ is $u^\mu \nabla_\mu u^\alpha =0$. There is an equivalence class of connections which describe the same geodesic paths. Any connection in this class is obtained from another by the
projective transformation, which can be given by the one-form $\xi^\mu$ as 
\beq \label{projective}
\Gamma^{\lambda}{}_{\mu\nu}\rightarrow \Gamma^{\lambda}{}_{\mu\nu}+\delta^{\lambda}_{\mu}\xi_{\nu}\,.
\eeq
In analogy with the fact that rescaling of magnitudes has a priori nothing to do with the connection, the projective transformation has a priori nothing to do with the metric. It depends completely on the geometric prescription of the physical theory, whether they are related and what the relation would be. When the one-form above is a gradient, $\xi_\mu=\phi_{,\mu}$, we obtain the special case of the linear vector distortion transformation
(\ref{distortion}) with $w_2=1$, $w_1=w_3=0$. This has precisely the form of the electromagnetic gauge transformation \cite{Goenner:2014mka,Janssen:2018exh} (up to the imaginary factor \cite{Koivisto:2018aip}). 
This suggests a special geometric relevance for this behaviour of the connection under rescalings, and indeed such will be clarified below in \ref{tangent}. 

To summarise, we have arrived at a possible relation between the transformation (\ref{projective}) for the connection and the rescaling (\ref{ct}) for the metric, the two independent geometric objects of metric-affine geometry (or, Palatini formalism). We shall discard the relation (\ref{weyl}) from further consideration, since this relation was justified by the restriction to Riemannian geometry, which is ad hoc from our perspective, and besides has been already extensively studied in the past hundred years. Instead, we shall explore the transformations of metric-affine geometry, taking into account systematically all the three logical possibilities, as follows. 
\begin{itemize}
\item Only the connection is transformed. This is called the {\bf projective transformation}:
\be
\hat{\Gamma}^{\lambda}{}_{\mu\nu}=\Gamma^{\lambda}{}_{\mu\nu}+\delta^{\lambda}_{\mu}\xi_{\nu}\,, \quad \hat{g}_{\mu\nu} = g_{\mu\nu}\,. \label{hat}
\ee
\item Only the metric is transformed. This is called the {\bf conformal transformation}:
\be
\bar{\Gamma}^{\lambda}{}_{\mu\nu}=\Gamma^{\lambda}{}_{\mu\nu}\,, \quad \bar{g}_{\mu\nu} = e^{2\phi}g_{\mu\nu}\,. \label{bar}
\ee
\item Both the metric and the connection are transformed. This is called the {\bf frame rescaling}:
\be
\tilde{\Gamma}^{\lambda}{}_{\mu\nu}=\Gamma^{\lambda}{}_{\mu\nu}+\delta^{\lambda}_{\mu}\partial_\nu\phi\,, \quad \tilde{g}_{\mu\nu} = e^{2\phi}g_{\mu\nu}\,. \label{tilde}
\ee
\end{itemize}
All of them can be interpreted as scale transformations, or, ``calibrations''. 
As above, we shall denote the transformed quantities in the three cases with a hat (proper calibration), with a bar (orthonormal calibration) and with a tilde (holonomic calibration), respectively. 
The nomenclature can be explained better, and the physical interpretation of the transformations can be further elaborated, with the help of a further set of indices. For this purpose only we'll make a very brief excursion into the tangent space in the following subsection, which the reader may choose to skip.

\subsection{Scale Transformation in Metric-Affine Gauge Geometry}
\label{tangent}

To consider the underlying symmetry transformations in gravity theory it is useful to introduce a local basis $\ie_a{}^\mu$, called the frame field, or the vielbein, which is a set of $n$ vector fields. 
The coframe field $\e^a{}_\mu$ is the dual  set of covectors which satisfy $\e^a{}_\nu\ie_a{}^\mu=\delta^\mu_\nu$, and $\e^a{}_\mu\ie_b{}^\mu=\delta^a_b$. 
We have the metric $\eta_{ab}$ which is convenient to consider in the Minkowski form. In terms of the frame field, the spacetime metric is given as
\be \label{metric}
g_{\mu\nu} = \eta_{ab}\e^{a}{}_\mu \e^{b}{}_\nu\,.
\ee
A fundamental field is then the connection one-form $\alpha^a{}_{b\mu}$, and we denote covariant derivative the associated with this connection by $D_\mu$.
The spacetime affine connection coefficients are given as
\be \label{arel}
\Gamma^\alpha{}_{\mu\nu} = \ie_a{}^\alpha D_\mu \e^a{}_\nu =  -\e^a{}_\nu D_\mu \ie_a{}^\alpha\,.
\ee 
This definition is often called ``the tetrad postulate''. Explicitly, the covariant derivative involves the connection $\alpha^a{}_{b \mu}$ as,
\ba
D_\mu \e^a{}_\nu & = & \e^a{}_{\nu,\mu} + \alpha^a{}_{b \mu}\e^b{}_\nu \\
D_\mu \ie_a{}^\nu & = & \ie_a{}^{\nu}{}_{,\mu} + \alpha_a{}^{b}{}_{\mu}\ie_b{}^\nu
\ea
and therefore the interpretation of the ``tetrad postulate'' is that the vielbein constant in the sense that it is preserved by the differentiation with respect to the sum of the two connections.
  
A general linear transformation is parameterised by $\Lambda^a{}_b$, which has an inverse
$\Lambda^a{}_c(\Lambda^{-1}){}^{c}{}_b= (\Lambda^{-1}){}^{a}{}_c\Lambda^c{}_b=\delta^a_b$,
and its action in the orthonormal geometry is
\begin{subequations}
\ba
\e^a{}_\mu & \rightarrow & \Lambda^a{}_b \e^b{}_\mu \\
\ie_a{}^\mu & \rightarrow & {(\Lambda^{-1})}{}^b{}_a \ie_b{}^\mu \\
\alpha^a{}_{b \mu} & \rightarrow & \Lambda^a{}_c\lp \alpha^c{}_{d\mu}  - \delta^c_d\partial_\mu\rp (\Lambda^{-1}){}^d{}_b \\
\eta_{ab} & \rightarrow & \eta_{ab}\,. 
\ea 
\end{subequations}
The trace of the general linear transformation can parameterised by one parameter $\phi$ as $\bar{\Lambda}^a{}_b=e^{\phi}\delta^a_b$. Explicitly, we have then 
\begin{subequations}
\label{orthonormal} 
\ba
\bar{\e}^a{}_\mu & = & e^{\phi} \e^a{}_\mu \\
\bar{\ie}_a{}^\mu & = & e^{-\phi} \ie_a{}^\mu \\
\bar{\alpha}^a{}_{b \mu} & = & {\alpha}^a{}_{b \mu} + \phi_{,\mu}\delta^a_b \\
\bar{\eta}_{ab} & = & \eta_{ab}
\ea
\end{subequations}
which gives a gauge transformation of the spin connection. From (\ref{metric}) and (\ref{arel}), 
we have then 
\ba
\bar{g}_{\mu\nu} & = & e^{2\phi}g_{\mu\nu} \label{htrans_t} \\
\bar{\Gamma}^\alpha{}_{\mu\beta} & = & \Gamma^\alpha{}_{\mu\beta} \,. \label{htrans_c}
\ea
Therefore, the trace of the general linear transformation in orthonormal geometry is manifested in spacetime geometry as what we call the conformal transformation (\ref{bar}) in this paper.

We can also consider a mere rescaling of the frame field, such that $\hat{\e}^a{}_\mu  =  e^{\phi} \e^a{}_\mu$,
$\hat{\alpha}^a{}_{b \mu}= {\alpha}^a{}_{b \mu}$, which in view of  (\ref{metric}) and (\ref{arel}) implies that
\ba
\tilde{g}_{\mu\nu} & = & e^{2\phi}g_{\mu\nu} \label{trans_t} \\
\tilde{\Gamma}^\alpha{}_{\mu\beta} & = & \Gamma^\alpha{}_{\mu\beta} + \phi_{,\mu}\delta^\alpha_\beta\,. \label{trans_c}
\ea
The latter transformation is a special case $\lambda_\mu=\phi_{,\mu}$ of what Einstein and others called the ``$\lambda$-transformation'' \cite{Goenner:2014mka}
\be
\hat{\Gamma}^\alpha{}_{\mu\beta} \rightarrow \Gamma^\alpha{}_{\mu\beta}  + \lambda_{\mu}\delta^\alpha_{\beta}\,. \label{lambda}
\ee
This transformation amounts to a reparameterisation of the geodesic parameters, and thus leaves the autoparallels of the connection invariant. 
The case is of course the opposite to the frame rescaling in the sense that we obtain it by considering that the tangent space connection transforms as $\alpha^a{}_{b \mu}  \rightarrow  \Lambda^a{}_c\lp \alpha^c{}_{d\mu}  - \delta^c_d\partial_\mu\rp (\Lambda^{-1}){}^d{}_b$, but that the frame does not transform, $\e^a{}_\mu \rightarrow \e^a{}_\mu$. 
Clearly, for these reasons we can regard the combined effect (\ref{trans_t},\ref{trans_c}) in spacetime as a rescaling of the frame and the shift (\ref{lambda}) of the spacetime affinity as the projection of the connection in the tangent space. 

In the orthonormal geometry we have considered up to this point, the tangent space can be seen to be defined by the constancy of the algebraic object $\eta_{ab}$. There is however another possible viewpoint. 
In a holonomic prescription, we rather regard the frame field as an absolute invariant. Then, it is not changed by any transformation, but rather transformations are understood to occur with respect  to the frame field. The price to pay is then, obviously, that $\eta_{ab}$ has to be allowed to become ``relative''. In this prescription, a general linear transformation is  
\begin{subequations}
\ba
\e^a{}_\mu & \rightarrow &  \e^b{}_\mu \\
\ie_a{}^\mu & \rightarrow &  \ie_b{}^\mu \\
\alpha^a{}_{b \mu} & \rightarrow & \Lambda^a{}_c\lp \alpha^c{}_{d\mu}  - \delta^c_d\partial_\mu\rp (\Lambda^{-1}){}^d{}_b \\
\eta_{ab} & \rightarrow & \eta_{cd}(\Lambda^{-1}){}^c{}_a(\Lambda^{-1}){}^d{}_b\ \,. 
\ea 
\end{subequations}
Now the trace of the general transformation, if parameterised as $\tilde{\Lambda}^a{}_b=e^{-\phi}\delta^a_b$, results in $\tilde{\eta}_{ab}=e^{2\phi}\eta_{ab}$ and thus, from (\ref{metric}) in that
$\tilde{g}_{\mu\nu} = e^{2\phi}g_{\mu\nu}$, and from (\ref{arel}) we readily obtain the same result as in (\ref{trans_c}). Thus, the trace of the linear transformation in holonomic geometry can be interpreted as a frame  
rescaling in orthogonal geometry. Alternatively we could thus (\ref{bar}) the ``orthogonal calibration'' and (\ref{tilde}) the ``holonomic calibration''. 

Finally, the projective transformation (\ref{hat}) could be called the ``proper calibration''. The spacetime metric (\ref{metric}) is obviously left invariant if we transform both the frame and the metric of the tangent space, i.e.
consider the general linear transformation of the tangent space variables
\begin{subequations}
\label{proper}
\ba
\e^a{}_\mu & \rightarrow &  \Lambda^a{}_b\e^b{}_\mu \\
\ie_a{}^\mu & \rightarrow &  {(\Lambda^{-1})}{}^b{}_a\ie_b{}^\mu \\
\alpha^a{}_{b \mu} & \rightarrow & \Lambda^a{}_c\lp \alpha^c{}_{d\mu}  - \delta^c_d\partial_\mu\rp (\Lambda^{-1}){}^d{}_b \\
\eta_{ab} & \rightarrow & \eta_{cd}(\Lambda^{-1}){}^c{}_a(\Lambda^{-1}){}^d{}_b\ \,.  
\ea 
\end{subequations}
The trace transformation should be now parameterised as $\tilde{\Lambda}^a{}_b=e^{\frac{1}{2}\phi}\delta^a_b$ to result exactly in (\ref{hat}). Thus, we can conclude that all the three cases can be interpreted as rescalings. The nomenclature we have adopted is justified from the orthonormal perspective\footnote{Needless to say, this terminology is not systematically used in the literature, but typically any version that is adopted is referred to as ``the conformal transformation''. However,
our definition of the conformal transformation (\ref{bar}) agrees with the review \cite{Hehl:1994ue}, since the transformation (\ref{orthonormal}) can be seen as the volume changing part of the proper linear transformation (\ref{proper}) (which is appropriately called ``the local scale transformation'' in \cite{Hehl:1994ue}) that is generalised  ``by admitting arbitrary exponents'' of the rescaling factor $e^{\phi}$, in the case of (\ref{orthonormal}) in particular giving the metric $\eta_{ab}$ the rescaling weight $2$. In the same way, the frame rescaling (\ref{tilde}) is simply the local scale transformation, i.e. the trace part of (\ref{proper}), accompanied with the non-trivial rescaling weight $1$ for the frame field and $-1$ for the coframe field. In the context of torsion transformations, the $\tilde{\delta}$-transformation (\ref{tilde}) and the $\bar{\delta}$-transformation (\ref{bar}) correspond to the ``strong conformal symmetry'' and the ``weak conformal symmetry'' \cite{Shapiro:2001rz}, respectively. }. 

In this paper our focus is on scale-invariance, but these considerations can be straightforwardly generalised for an arbitrary (linear) transformation.

\section{A Simple Scale-Invariant Theory}
\label{simple}

As a warm-up, let us study now a conformally invariant theory by coupling the Ricci scalar $R$ to a scalar field $\psi$, in the metric-affine framework. The nice thing now is that one does not need the existence of an additional gauge field $A_{\mu}$ in order to define the gauge covariant derivative on $\psi$ since torsion and non-metricity offer enough room to accommodate it into them. 
To be more specific, consider the action
\beq
S=\frac{1}{2\kappa}\int \diff^{n}x \Big[\sqrt{-g}\psi^{2}R+\lambda \sqrt{-g} g^{\mu\nu}D_{\mu}\psi D_{\nu}\psi \Big] \label{confthe}
\eeq
where $\lambda$ is a parameter and $D_{\mu}\psi$ the gauge covariant derivative on the field, to be defined in a moment. Note that the Dirac-Weyl theory \cite{Dirac:1973gk,israelit} corresponds to the above action
where the metric-affine $R$ is replaced by the Einstein-Hilbert scalar, the $\lambda$ is fixed to the special value of the conformal coupling, and an additional gauge field $A_\mu$ is introduced to define the covariant
derivative $D_\mu$. 

Notice now that the first term in the above action is invariant under conformal transformations of the metric
\beq
g_{\mu\nu} \rightarrow \bar{g}_{\mu\nu}=e^{2 \phi}g_{\mu\nu}
\eeq
provided that we simultaneously transform the scalar field as
\beq
\psi \rightarrow \bar{\psi}=e^{\frac{(2-n)}{2}\phi}\psi\,.
\eeq
In order to keep this invariance on the kinetic term too, one needs to replace the partial derivative $\partial_{\mu}$ with a covariant one $D_{\mu}=\partial_{\mu}+A_{\mu}$ and  also impose a gauge transformation on the field $A_{\mu}$ ($A_{\mu}\rightarrow A_{\mu}+\partial_{\mu}\phi$) so as to have the transformation
\beq
\bar{D}_{\mu}\bar{\psi}=e^{\frac{(2-n)}{2}\phi}D_{\mu}\psi
\eeq
and subsequently
\beq
\sqrt{-\bar{g}}\bar{g}^{\mu\nu}\bar{D}_{\mu}\bar{\psi}\bar{D}_{\nu}\bar{\psi}=g^{\mu\nu}D_{\mu}\psi D_{\nu}\psi \label{Deq}
\eeq
which will ensure the conformal invariance of the total action. Now, what's interesting is that we do not have to add this gauge field $A_{\mu}$ by hand, we have a generalized geometry offering torsion and non-metricity vectors that can do the job. Notice now that since the torsion vector $S_{\mu}$ does not change under conformal transformations, it cannot be regarded as our desired gauge field. The non-metricity (Weyl) vector however, transforms as
\beq
\bar{Q}_{\mu}=Q_{\mu}-2 n \partial_{\mu} \phi
\eeq
under a conformal transformation. Therefore, defining the covariant derivative on the scalar field as
\beq
D_{\mu}\equiv \partial_{\mu}+\frac{2-n}{4n}Q_{\mu}
\eeq
ensures that (\ref{Deq}) is satisfied. So, building the action this way, let us derive the field equations of ($\ref{confthe}$). Variation with respect to the metric tensor yields
\begin{gather}
-\frac{1}{2}g_{\mu\nu}\Big( \psi^{2}R +\lambda (D\psi)^{2}\Big)+\psi^{2}R_{(\mu\nu)}+\lambda D_{\mu}\psi D_{\nu}\psi   
+ \lambda \frac{(n-2)}{2 n}g_{\mu\nu} \frac{\partial_{\alpha}(\sqrt{-g}\psi D^{\alpha} \psi)}{\sqrt{-g}}=0
\end{gather}
where we have abbreviated $(D\psi)^{2}=g^{\mu\nu}D_{\mu}\psi D_{\nu}\psi$. Now, since our initial action is conformally invariant one would expect that the trace of the above equation identically vanishes. In fact, the trace of the above equation gives the same equation that one gets when varying with respect to the scalar field $\psi$. Therefore, when the equation of motion for $\psi$ is on shell, the above trace vanishes identically. To see this first note that the trace of the above field equations is
\beq
\psi^{2}R +\lambda (D\psi)^{2}-\lambda \frac{\partial_{\alpha}(\sqrt{-g}\psi D^{\alpha} \psi)}{\sqrt{-g}} =0\,. \label{Rpsi}
\eeq
On the other hand, varying the action with respect to $\psi$, we obtain
\beq
R \psi -\lambda \frac{\partial_{\alpha}(\sqrt{-g} D^{\alpha} \psi)}{\sqrt{-g}}-\lambda \frac{(n-2)}{4 n}Q^{\mu}(D_{\mu}\psi)=0\,.
\eeq
Multiplying this by $\psi$ (given that $\psi \neq 0$) and doing a partial integration it follows that
\beq
R \psi^{2}-\lambda \frac{\partial_{\alpha}(\sqrt{-g}\psi D^{\alpha} \psi)}{\sqrt{-g}}+\lambda \Big(\partial_{\mu}+\frac{2-n}{4n}Q_{\mu}\Big)D^{\mu}\psi =0
\eeq
or equivalently 
\beq
\psi^{2}R +\lambda (D\psi)^{2}-\lambda \frac{\partial_{\alpha}(\sqrt{-g}\psi D^{\alpha} \psi)}{\sqrt{-g}} =0
\eeq 
which is indeed the same equation with ($\ref{Rpsi}$). Lastly, variation of the action with respect to the connection yields
\beq
P_{\lambda}{}^{\mu\nu}(h)+\lambda\frac{(2-n)}{n}\delta_{\lambda}^{\mu}(D^{\nu}\psi)=0 \label{Palahx}
\eeq
where
\begin{gather}
P_{\lambda}{}^{\mu\nu}(h) \equiv -\frac{\nabla_{\lambda}(\sqrt{-g}\psi^{2}g^{\mu\nu})}{\sqrt{-g}}+\frac{\nabla_{\alpha}(\sqrt{-g}\psi^{2}g^{\mu\alpha}\delta_{\lambda}^{\nu})}{\sqrt{-g}} +
2 \psi^{2}(S_{\lambda}g^{\mu\nu}-S^{\mu}\delta_{\lambda}^{\nu}-  S_{\lambda}{}^{\mu\nu}) 
\end{gather}
is the Palatini tensor computed with respect to the metric $h_{\mu\nu}=\psi^{2} g_{\mu\nu}$. This tensor can also be written as
\beq
P_{\lambda}{}^{\mu\nu}(h)=\psi^{2}P_{\lambda}{}^{\mu\nu}(g)+\delta_{\lambda}^{\nu}g^{\mu\alpha}\partial_{\alpha}\psi^{2}-g^{\mu\nu}\partial_{\lambda}\psi^{2}
\eeq
where $P_{\lambda}{}^{\mu\nu}(g)$ is the usual Palatini tensor computed with respect to the metric tensor $g_{\mu\nu}$. Looking back at ($\ref{Palahx}$), contracting in $\mu=\lambda$ and using the fact that the Palatini tensor is traceless in its first two indices\footnote{Note that both $P_{\mu}{}^{\mu\nu}(g)=0$ and $P_{\mu}{}^{\mu\nu}(h)=0$, that is any Palatini tensor that is built from a metric conformally related to $g_{\mu\nu}$ is also traceless in its first two indices. We shortly return to study this systematically in Section \ref{identities}.}, it follows that
\beq
D^{\nu}\psi=0 \label{psieq}
\eeq
which when substituted back at ($\ref{Palahx}$) gives
\begin{gather}
P_{\lambda}{}^{\mu\nu}(h) =0 \quad \Rightarrow \quad
\psi^{2}P_{\lambda}{}^{\mu\nu}(g)=-\delta_{\lambda}^{\nu}g^{\mu\alpha}\partial_{\alpha}\psi^{2}+g^{\mu\nu}\partial_{\lambda}\psi^{2}\,.
\end{gather}
With this at hand we can use the connection decomposition and easily find the affine connection
\beq
\Gamma^{\lambda}{}{}_{\mu\nu}=\tilde{\Gamma}{}{}_{\mu\nu}+\frac{2}{n-2}g_{\mu\nu}\frac{\partial^{\lambda}\psi}{\psi}-\frac{2}{n-2}\delta^{\lambda}_{\nu}\frac{\partial_{\mu}\psi}{\psi}+\frac{1}{2}\delta^{\lambda}_{\mu}\tilde{Q}_{\nu}\,.
\eeq
Before finding the expressions for torsion and non-metricity that follow from the above, let us expand ($\ref{psieq}$) to get
\beq
\partial_{\mu}\psi -\frac{(n-2)}{4n}Q_{\mu} \psi =0
\quad \Rightarrow \quad Q_{\mu}=\frac{4 n}{n-2}\frac{\partial_{\mu}\psi}{\psi}
\eeq
that is, the Weyl vector is exact and powered by the scalar field $\psi$. Now, using the above connection decomposition and the fact that
\beq
S_{\mu\nu}{}^{\lambda}=N^{\lambda}{}{}_{[\mu\nu]}\,\,\,\,
\text{and} \,\,\,\,
Q_{\alpha\mu\nu}=2 N_{(\alpha\mu)\nu}\,, \,\,\,\, \text{where} \,\,\,\, N^{\lambda}{}{}_{\mu\nu} \equiv \Gamma^{\lambda}{}{}_{\mu\nu}-\left\{^{\phantom{i} \lambda}_{\mu\nu}\right\}
\eeq
it follows that
\beq
S_{\mu\nu}{}^{\lambda}=-2\frac{\partial_{[\mu}\psi \delta^{\lambda}_{\nu]}}{\psi}+\frac{1}{2}\delta^{\lambda}_{\mu} \tilde{Q}_{\nu]}\,\,\,\,
\text{and} \,\,\,\,
Q_{\alpha\mu\nu}=\tilde{Q}_{\alpha}g_{\mu\nu} = \frac{1}{n} Q_{\alpha}g_{\mu\nu}
\eeq
where the last equality is deduced by the contraction of the previous one. 
Also recalling that $Q_{\mu}=\frac{4 n}{n-2}\frac{\partial_{\mu}\psi}{\psi}$ we have
\beq
Q_{\alpha\mu\nu}=\frac{4}{n-2}g_{\mu\nu}\frac{\partial_{\mu}\psi}{\psi}
\eeq
which is the case of a Weyl integrable non-metricity. Also, using the above, the torsion tensor may be expressed as
\beq
S_{\mu\nu}{}^{\lambda}=\frac{4}{n-2}\frac{\delta_{[\mu}^{\lambda}\partial_{\nu]}\psi}{\psi}
\eeq
with the torsion vector
\beq
S_{\mu}=-\frac{2(n-1)}{(n-2)}\frac{\partial_{\mu}\psi}{\psi}
\eeq
and the above is a case of vectorial torsion with an exact torsion vector. We note at this point that the torsion and the non-metricity are dual to each. This is because they only appear in their vectorial forms in the
theory under consideration (and further, pure gauge vectors), which the projective transformations change into each other, and the curvature is projectively invariant. Therefore, from the above construction of the
conformally invariant theory, we get the corresponding frame rescaling invariant theory by prescribing the gauge field $A_\mu$ defining the covariant derivative of the scalar to be given by torsion instead of non-metricity. 
To summarise, the three cases of symmetry correspond to the following three prescriptions for the covariant derivative of the scalar field in the action (\ref{confthe}). 
\begin{itemize}
\item projective invariance:
\begin{subequations}
\label{gcov}
\be
D_{\mu} \equiv \nabla_{\mu} \quad \Rightarrow \quad D_\mu\psi = \partial_\mu\psi\,. \label{gcovP}
\ee
\item conformal invariance: 
\be 
D_{\mu}\equiv \nabla_{\mu}-\bar{\mathrm{w}}\left(\frac{n-2}{4n}\right)Q_{\mu}  \quad \Rightarrow \quad
D_{\mu}\psi = \lb\partial_{\mu}-\left(\frac{n-2}{4n}\right)Q_{\mu}\rb\psi \label{gcovQ} 
\ee
\item frame rescaling invariance:
\be
D_{\mu}  \equiv \nabla_{\mu}-\tilde{\mathrm{w}}\left(\frac{n-2}{n-1}\right)S_{\mu} \quad \Rightarrow \quad
D_{\mu}\psi  \equiv \lb \partial_{\mu}-\left(\frac{n-2}{n-1}\right)S_{\mu}\rb\psi \label{gcovS}
\ee
\end{subequations}
\end{itemize}
In each case, we have written the definition that applies for an arbitrary tensor with the corresponding weight given by the $\mathrm{w}$-symbol, and then specified the action on the rank-0 tensor
$\psi$.
For the corresponding projectively invariant theory, the compensating scalar field is not needed at all (if it is included, its transformation should be considered trivial to retain the projective invariance). The duality of torsion and non-metricity vectors in a projectively invariant theory was elaborated in great detail in the recent Ref. \cite{Iosifidis:2018zjj}. 

Finally, using the above results, the field equations for the scalar field and the metric imply the field equations
\beq
R=0 \quad \text{and} \quad
R_{\mu\nu}=0\,.
\eeq
We can now clarify the physical interpretation of the theory (\ref{confthe}). It is equivalent to Einstein's Gravity in vacuum, even though the curvature has more degrees of freedom coming from torsion and non-metricity. In this simple conformally invariant model we have a Weyl non-metricity and vectorial torsion both sourced by the scalar field $\psi$, but their role is that of a pure-gauge field. A natural gauge fixing is to choose the transformation parameter $\phi$ such that $\psi=1$ is a constant. This could be seen as the unitary gauge, where both torsion and non-metricity then vanish, and we recover GR with the correct normalisation of the gravitational constant $\kappa$. In any other gauge the gravitational coupling would appear to be a function of time and space, effectively $\kappa \rightarrow \kappa/\psi$, but the physics of the theory would be rendered equivalent by the dynamics of the gauge fields $S_\mu$ and $Q_\mu$. The equivalence would be broken by adding kinetic terms for these fields, corresponding to the scalars we have listed in the Appendix \ref{scalars}.
Let us now proceed to the study of those scalars.    

\section{The Parity-Even Quadratic Action}
\label{quadratic}

We are interested in quadratic, second order metric-affine theories which are covariant under the three scaling transformations. Let us first note the fact that any curvature scalar\footnote{There is 1 linear and 10 quadratic scalars \cite{Jimenez:2014rna} in the parity-even case, and in total there is an infinite number of such scalars.} that may be constructed in the Palatini formalism is covariant under conformal transformation and invariant under the pure-gauge projective transformation. Trivially, these curvature scalars are then covariant under the frame transformation as well. Therefore, after the simple example with curvature in Section \ref{simple}, we shall mostly focus on actions that are quadratic in torsion and non-metricity. Another rationale for our choice of action is that it is the most general quadratic theory involves no derivatives of the connection but is up to second order in derivatives of the metric \cite{Pagani:2015ema}. 

\subsection{The Scale-Covariant Scalars}

In this Section consider the parity-symmetric action in arbitrary spacetime dimension $n$. It is given as 
\beq
S=\frac{1}{2 \kappa}\int \diff^{n}x \sqrt{-g} \Big[ \mathcal{L}^{+}_{Q}+ \mathcal{L}^{+}_{T}+ \mathcal{L}^{+}_{QT} \Big] +S_{Matter} \label{plus}
\eeq
which is parameterised by the $5+3+3=11$ parameters $a_i$, $b_i$ and $c_i$ as follows:
\begin{subequations}
\label{even_s}
\ba
\mathcal{L}_{Q}^{+} & = & a_{1}Q_{\alpha\mu\nu}Q^{\alpha\mu\nu} +
a_{2}Q_{\alpha\mu\nu}Q^{\mu\nu\alpha} +
a_{3}Q_{\mu}Q^{\mu}+
a_{4}q_{\mu}q^{\mu}+
a_{5}Q_{\mu}q^{\mu}  \\
 \mathcal{L}_{T}^{+} & = & b_{1}S_{\alpha\mu\nu}S^{\alpha\mu\nu} +
b_{2}S_{\alpha\mu\nu}S^{\mu\nu\alpha} +
b_{3}S_{\mu}S^{\mu} \\
  \mathcal{L}_{QT}^{+} & = & c_{1}Q_{\alpha\mu\nu}S^{\alpha\mu\nu}+
c_{2}Q_{\mu}S^{\mu} +
c_{3}q_{\mu}S^{\mu}\,. 
\ea
\end{subequations}
For the systematical deduction of this action, we refer the reader to the Appendix \ref{scalars}. 

To begin with, let us consider how this action can be considered as a limit to GR. We could of course add the Einstein-Hilbert Lagrangian $\mathcal{L}_{EH} = \mathcal{R}-2\Lambda$ to the above and consider the quadratic terms as post-Riemannian corrections. Notice now however that for the parameter choice $b_{1}=1$, $b_{2}=-2$, $b_{3}=-4$, $a_{i}=0=c_{i}$  one recovers the teleparallel equivalent of GR by imposing a vanishing curvature and non-metricity, as it was shown recently in Refs. \cite{BeltranJimenez:2017tkd,BeltranJimenez:2018vdo}. It was further shown there that by demanding vanishing curvature and torsion and taking $a_{1}=-a_{3}=1/4$, $a_{2}=-a_{5}=-1/2$, $a_{4}=0$, $b_{i}=0=c_{i}$ one obtains the symmetric teleparallel equivalent of GR \cite{Nester:1998mp,Adak:2008gd} from the above action. Furthermore if we pick $b_{1}=1$, $b_{2}=-1$, $b_{3}=-4$ , $a_{1}=-a_{3}=1/4$, $a_{2}=-a_{5}=-1/2$, $a_{4}=0$,  $c_{1}=-c_{2}=c_{3}=2$ and impose only the vanishing of curvature, we may expect to reproduce a generalized equivalent to GR that admits both torsion and non-metricity. The latter possibility has not, however, been considered in detail previously. 

Now, in order to obtain a conformally invariant theory we should first restrict the above parameters and find a specific combination for which the total Lagrangian density transforms covariantly under the conformal transformation (\ref{bar}), namely it only picks up a factor $e^{- 2 \phi}$. To do so, we use the transformation laws for the quadratic scalars that are derived in detail in the Appendix \ref{b_bar}. Then under a conformal transformation, we have
\begin{subequations}
\ba
\bar{\mathcal{L}}^+_{T} & = & e^{-2\phi}\mathcal{L}^+_{T} \\
\bar{\mathcal{L}}^+_{Q} & = & e^{-2\phi}\mathcal{L}^+_{Q}-e^{-2\phi}Q^{\mu}\partial_{\mu}\phi (4a_{1}+4 n a_{3}+2 a_{5}) 
-e^{-2\phi}q^{\mu}\partial_{\mu}\phi ( 4 a_{2}+4 a_{4}+2 n a_{5}) \nonumber \\
& + & e^{-2\phi}(\partial \phi)^{2}4( n a_{1}+a_{2}+ n^{2} a_{3}+a_{4}+n a_{5}) \\
\bar{\mathcal{L}}^+_{QT} & = & e^{-2\phi}\mathcal{L}^+_{QT}-e^{-2\phi}2 S^{\mu}\partial_{\mu}\phi (  c_{1} +n c_{2}+c_{3})\,.
\ea
\end{subequations}
From these we conclude that the parameter choices fo $a_i$ and $c_i$ for which 
\begin{subequations}
\label{confinv}
\ba
 0 & = & 4a_{1}+4 n a_{3}+2 a_{5}  \\  0 & = & 4 a_{2}+4 a_{4}+2 n a_{5} \\  0 & = & n a_{1}+a_{2}+ n^{2} a_{3}+a_{4}+n a_{5}  \\
  0 & = & c_{1}+n c_{2}+c_{3} 
\ea
\end{subequations}
and whatever $b_{i}'s$ guarantee that
\beq
\bar{\mathcal{L}}^+_{Q}+\bar{\mathcal{L}}^+_{T}+\bar{\mathcal{L}}^+_{QT}=e^{-2\phi}\Big(\mathcal{L}^+_{Q}+ \mathcal{L}^+_{T}+\mathcal{L}^+_{QT} \Big)
\eeq
as we desired. Since there are 4 constraints on the parameters, we see that the most general conformally covariant quadratic action is given by 11-4=7 free parameters. 

Now, let us consider frame rescalings (\ref{tilde}). Some details of the derivation are found in the Appendix \ref{b_tilde}, the result being 
\begin{subequations}
\ba
\tilde{\mathcal{L}}^+_{Q} & = &  e^{-2 \phi}\mathcal{L}^+_{Q} \\
\tilde{\mathcal{L}}^+ _{T}  & = &  e^{-2 \phi}\mathcal{L}^+_{T}-e^{-2 \phi}S^{\mu}\partial_{\mu}\phi \Big( 2 b_{1}-b_{2}+(n-1)b_{3} \Big) 
+\frac{(n-1)}{4} e^{-2 \phi}(\partial \phi)^{2}\Big( 2 b_{1} -b_{2}+(n-1)b_{3} \Big) \\
\tilde{\mathcal{L}}^+_{QT} & = &  e^{-2 \phi}\mathcal{L}^+_{QT} - \frac{1}{2}e^{-2 \phi}Q^{\mu}\partial_{\mu}\phi \Big( c_{1}+(n-1)c_{2} \Big) 
+\frac{1}{2} e^{-2 \phi}q^{\mu}\partial_{\mu}\phi  \Big(c_{1}+(1-n)c_{3} \Big)\,.
\ea
\end{subequations}
Then, frame rescaling invariance 
\beq
\tilde{\mathcal{L}}^+_{Q}+\tilde{\mathcal{L}}^+_{T}+\tilde{\mathcal{L}}^+_{QT}=e^{-2\phi}\Big(\mathcal{L}^+_{Q}+ \mathcal{L}^+_{T}+\mathcal{L}^+_{QT} \Big)
\eeq
is ensured so long we have $b_i$'s and $c_i$'s that satisfy
\begin{subequations}
 \label{frinv}
\ba
0 & = &  2 b_{1}-b_{2}+(n-1)b_{3} \\
0 & = &   c_{1}+(n-1) c_{2}=0 \\
0 & = &  c_{1}-(n-1) c_{3}=0
\ea
\end{subequations}
and whatever $a_{i}'s$. There are thus 8 independent $\tilde{\delta}$-covariant combinations of scalars.

Now, let us see how our action changes under projective transformations of the connection (\ref{hat}),
$\Gamma^{\lambda}{}_{\mu\nu}\longrightarrow \hat{\Gamma}^{\lambda}{}_{\mu\nu} =\Gamma^{\lambda}{}_{\mu\nu}+ \delta_{\mu}^{\lambda}\xi_{\nu}$
which do not affect the spacetime metric
$g_{\mu\nu}\longrightarrow \hat{g}_{\mu\nu}=g_{\mu\nu}$. Note that we do not require the vector $\xi^\mu$ to be a gradient.
We compute
\begin{subequations}
\ba
\hat{\mathcal{L}}^+_{Q} & = & \mathcal{L}^+_{Q}+(4 a_{1}+4 n a_{3}+2 a_{5})Q_{\mu}\xi^{\mu}+(4 a_{2}+4 a_{4}+2 n a_{5})q_{\mu}\xi^{\mu}
+(4 n a_{1}+4 a_{2}+4 n^{2} a_{3}+4 a_{4}+4 n a_{5})\xi_{\mu}\xi^{\mu}\,\,\, \\
\hat{\mathcal{L}}^+_{T} & = & \mathcal{L}^+_{T}-\lb 2 b_{1}-b_{2}+(n-1) b_{3}  \rb S_{\mu}\xi^{\mu}
+\frac{(n-1)}{4}\lb 2 b_{1}-b_{2}+(n-1) b_{3} \rb \xi_{\mu}\xi^{\mu} \\
\hat{\mathcal{L}}^+_{QT} & = & \mathcal{L}^+_{QT}-\frac{1}{2}\lb c_{1}+(n-1)c_{2} \rb Q_{\mu}\xi^{\mu}+\frac{1}{2}\lb c_{1}-(n-1)c_{3} \rb q_{\mu}\xi^{\mu} 
+ (c_{1}+n c_{2}+ c_{3})\lb 2S_{\mu}\xi^{\mu} - (n-1)\xi_{\mu}\xi^{\mu}\rb\,.
\ea
\end{subequations}
Therefore, the total action changes according to
\ba
\hat{\mathcal{L}}^+_{Q}+\hat{\mathcal{L}}^+_{T}+\hat{\mathcal{L}}^+_{QT} & = & \mathcal{L}_{Q}^++\mathcal{L}^+_{T}+ \mathcal{L}^+_{QT} \nonumber \\
& + & \left[ 2(2 a_{1}+2 n a_{3}+a_{5})-\frac{1}{2}\Big( c_{1}+(n-1)c_{2} \Big) \right]Q_{\mu}\xi^{\mu} \nonumber \\
& + & \left[ 2(2 a_{2}+2  a_{4}+n a_{5})+\frac{1}{2}\Big( c_{1}-(n-1)c_{3} \Big) \right]q_{\mu}\xi^{\mu} \nonumber \\
& + & \Big[ -2 b_{1}+b_{2}-(n-1) b_{3}+2(c_{1} +n c_{2}+c_{3}) \Big]S_{\mu}\xi^{\mu} \nonumber \\
& + & \left[4\lp n a_{1}+ a_{2}+ n^{2} a_{3}+ a_{4}+ n a_{5}\rp +\frac{(n-1)}{4}\Big( 2 b_{1}-b_{2}+(n-1) b_{3}- 4(c_{1}+n c_{2}+c_{3}) \Big) 
  \right]\xi_{\mu}\xi^{\mu}\,. \quad
\ea
Then, projective invariance is ensured if the parameters satisfy
\begin{subequations}
\label{proinv}
\ba
 0 & = &4 (2 a_{1}+2 n a_{3}+a_{5})-c_{1}-(n-1)c_{2}
\\
 0 & = &4 (2 a_{2}+2  a_{4}+n a_{5})+c_{1}-(n-1)c_{3}
\\
 0 & = &2 b_{1}-b_{2}+(n-1) b_{3}-2(c_{1} +n c_{2}+c_{3})
\\
 0 & = &16 (n a_{1}+ a_{2}+ n^{2} a_{3}+ a_{4}+ n a_{5})
+(n-1)\Big( 2 b_{1}-b_{2}+(n-1) b_{3}-4 (c_{1}+n c_{2}+c_{3})\Big)\,.
\ea
\end{subequations}
The important thing to note here is that the parameters $a_{i},b_{i},c_{i}$ mix when one demands projective invariance.  This means that $\mathcal{L}^+_{Q},\,\,\mathcal{L}^+_{T}$ and $ \mathcal{L}^+_{QT}$ may not independently projective invariant though their sum is. This was not the case when we considered conformal and frame rescaling transformations where the parameters did not mix and  $\mathcal{L}^+_{Q},\,\mathcal{L}^+_{T}$ and $ \mathcal{L}^+_{QT}$ were all independently invariant under the associated transformations iff their sum was.

\subsection{The Field Equations}

Having restricted the parameter space in the each of the three cases we can now obtain an invariant theory by coupling the above to $\psi^{2}$. We first combine the case of conformal and frame rescaling transformations in a single action given by
\be
S = \frac{1}{2 \kappa}\int \diff^{n}x \sqrt{-g}\Big[ \psi^{2}\lp \mathcal{L}^+_{Q} +  \mathcal{L}^+_{QT} + \mathcal{L}^+_{T}\rp +\lambda g^{\mu\nu}D_{\mu}\psi D_{\nu}\psi\Big] 
\ee
where again $\lambda$ is a parameter, $D_{\mu}$ is the gauge covariant derivative to be defined later, and $\mathcal{L}^+_{Q}+\mathcal{L}^+_{T}+ \mathcal{L}^+_{QT}$ was specified in (\ref{even_s}).
Now, it will be convenient for the calculations to define the ``superpotentials''
\begin{subequations}
\ba
\Omega^{\alpha\mu\nu} & \equiv & a_{1}Q^{\alpha\mu\nu}+a_{2} Q^{\mu\nu\alpha}+a_{3} g^{\mu\nu}Q^{\alpha}+a_{4}g^{\alpha\mu}q^{\nu}+a_{5}g^{\alpha\mu}Q^{\nu} \\
\Sigma^{\alpha\mu\nu} & \equiv & b_{1}S^{\alpha\mu\nu}+b_{2}S^{\mu\nu\alpha}+b_{3}g^{\mu\nu}S^{\alpha} \\
\Pi^{\alpha\mu\nu} & \equiv & c_{1}S^{\alpha\mu\nu}+c_{2}g^{\mu\nu}S^{\alpha}+c_{3}g^{\alpha\mu}S^{\nu}
\ea
\end{subequations}
for non-metricity, torsion and their mixing, respectively. With these, the above are written as
\begin{subequations}
\ba
\mathcal{L}^+_{Q} & = & Q_{\alpha\mu\nu}\Omega^{\alpha\mu\nu} \\ 
\mathcal{L}^+_{T} & = & S_{\alpha\mu\nu}\Sigma^{\alpha\mu\nu} \\
\mathcal{L}^+_{QT} & = & Q_{\alpha\mu\nu}\Pi^{\alpha\mu\nu}\,. 
\ea
\end{subequations}
We are now in a position to derive the variations of the above. Some intermediate steps in the derivations are confined to the Appendix \ref{identities}. Let us first compute variations with respect to the metric. We have
\begin{gather}
\sqrt{-g}\psi^{2}\delta_{g}\mathcal{L}^+_{Q}=(\delta g^{\mu\nu}) \Big[ \sqrt{-g}\psi^{2} L_{(\mu\nu)}+ (2S_{\lambda}-\nabla_{\lambda})J^{\lambda}{}_{\;(\mu\nu)}+g_{\mu\nu}(2S_{\lambda}-\nabla_{\lambda})\zeta^{\lambda} 
+\alpha_{4}(2S_{(\mu}-\nabla_{(\mu})( \sqrt{-g}\psi^{2}q_{\nu)}) \Big]
\end{gather}
where
\ba
L_{\mu\nu} & \equiv  & (a_{1}Q_{\mu\alpha\beta}+a_{2}Q_{\alpha\beta\mu})Q_{\nu}{}^{\alpha\beta} +  (a_{3}Q_{\mu}+a_{5}q_{\mu})Q_{\nu}+a_{3}Q_{\alpha\mu\nu}Q^{\alpha} \nonumber \\
&+ & Q_{\mu\nu\alpha}(a_{4}q^{\alpha}+a_{5}Q^{\alpha})-\Omega^{\alpha\beta}{}{}_{\nu}Q_{\alpha\beta\mu}-\Omega_{\alpha\mu\beta}Q^{\alpha\beta}{}{}_{\nu}
\ea
and we have also defined the tensor densities
\beq
J^{\lambda}{}_{\;(\mu\nu)} \equiv \sqrt{-g}\psi^{2}( \alpha_{1}Q^{\lambda}{}_{\mu\nu}+a_{2}Q_{\mu\nu}{}^{\lambda}+\Omega^{\lambda}{}_{\mu\nu})
\eeq
\beq
\zeta^{\lambda} \equiv \sqrt{-g}\psi^{2}(a_{3}Q^{\lambda}+a_{5}q^{\lambda})\,.
\eeq
Continuing with the  pure torsion and mixed part, we obtain
\beq
\sqrt{-g}\psi^{2}\delta_{g}\mathcal{L}^+_{T}=(\delta g^{\mu\nu}) \sqrt{-g}\psi^{2}\Big[ b_{1}(2S_{\nu\alpha\beta}S_{\mu}{}^{\alpha\beta}-S_{\alpha\beta\mu}S^{\alpha\beta}{}{}_{\nu})-b_{2}S_{\nu\alpha\beta}S_{\mu}{}^{\alpha\beta}+b_{3}S_{\mu}S_{\nu} \Big]
\eeq
and
\ba
\sqrt{-g}\psi^{2}\delta_{g}\mathcal{L}^+_{QT} & = & (\delta g^{\mu\nu}) \sqrt{-g}\psi^{2}\Big[\Pi_{\mu\alpha\beta}Q_{\nu}{}^{\alpha\beta}\nonumber \\
& - & ( c_{1}S_{\alpha\beta\nu}Q^{\alpha\beta}{}{}_{\mu}+c_{2}S^{\alpha}Q_{\alpha\mu\nu}+c_{3}S^{\alpha}Q_{\mu\nu\alpha}) 
+\frac{1}{\sqrt{-g}\psi^{2}}(2S_{\lambda}-\nabla_{\lambda})(\sqrt{-g}\psi^{2}\Pi^{\lambda}{}_{\mu\nu}) \Big]\,.
\ea
Using all the above we can now derive the field equations for the conformally  and frame rescaling invariant theories. 
To obtain a conformally invariant theory, the parameters must satisfy ($\ref{confinv}$) and the gauge covariant derivative on the scalar field has to be defined a in (\ref{gcovQ}).
On the other hand, in order to obtain a frame rescaling invariant theory, the parameter space is restricted by the constraints ($\ref{frinv}$) and the gauge covariant derivative should be defined as in (\ref{gcovS}).

Having clarified this, the field equations after varying with respect to the metric tensor are
\beq
\psi^{2}\Big( Z_{(\mu\nu)}-\frac{1}{2}g_{\mu\nu}\mathcal{L}\Big)-\frac{1}{2}g_{\mu\nu}\lambda (D\psi)^{2}+ \lambda \Big( D_{\mu}\psi D_{\nu}\psi +K_{\mu\nu}\Big)=0 \label{g_eom}
\eeq
where 
\ba
Z_{\mu\nu} & \equiv & L_{\mu\nu}+\xi_{\mu\nu}+b_{1}(2S_{\nu\alpha\beta}S_{\mu}{}^{\alpha\beta}-S_{\alpha\beta\mu}S^{\alpha\beta}{}{}_{\nu})-b_{2}S_{\nu\alpha\beta}S_{\mu}{}^{\alpha\beta}+b_{3}S_{\mu}S_{\nu} 
+\Pi_{\mu\alpha\beta}Q_{\nu}{}^{\alpha\beta} \nonumber \\ & - & ( c_{1}S_{\alpha\beta\nu}Q^{\alpha\beta}{}{}_{\mu}+c_{2}S^{\alpha}Q_{\alpha\mu\nu}+c_{3}S^{\alpha}Q_{\mu\nu\alpha}) 
 + \frac{1}{\sqrt{-g}\psi^{2}}(2S_{\lambda}-\nabla_{\lambda})(\sqrt{-g}\psi^{2}\Pi^{\lambda}{}_{\mu\nu}) 
 \ea
 and
 \be
\xi_{\mu\nu}  \equiv  \frac{1}{\sqrt{-g}\psi^{2}}\Big[ (2S_{\lambda}-\nabla_{\lambda})J^{\lambda}{}_{\;(\mu\nu)}+g_{\mu\nu}(2S_{\lambda}-\nabla_{\lambda})\zeta^{\lambda}
+\alpha_{4}(2S_{(\mu}-\nabla_{(\mu})( \sqrt{-g}\psi^{2}q_{\nu)}) \Big]\,.
\ee
Since 
\beq
\int \diff^{n}x\sqrt{-g} K_{\mu\nu}\equiv \int \diff^{n}x \sqrt{-g}(D^{\alpha}\psi)\frac{\delta(D_{\alpha}\psi)}{\delta g^{\mu\nu}}
\eeq
we have therefore
\begin{subequations}
\beq
K_{\mu\nu}  =  \frac{(n-2)}{2 n}g_{\mu\nu} \frac{\partial_{\alpha}(\sqrt{-g}\psi D^{\alpha} \psi)}{\sqrt{-g}} \label{k_bar}
\eeq
for the conformally invariant theory and
\beq
K_{\mu\nu} = 0
\eeq
\end{subequations}
for the frame rescaling invariant theory\footnote{This is so because in this case the gauge covariant derivative is constructed in terms of $S_{\mu}$ and the latter is independent of the metric tensor.}. 

Let us continue with the rest of the field equations. Variation with respect to the connection gives
\begin{gather}
\psi^{2} \Big( H^{\mu\nu}{}{}_{\lambda}+\delta^{\mu}_{\lambda}k^{\nu}+\delta^{\nu}_{\lambda}h^{\mu}+g^{\mu\nu}h_{\lambda}+f^{[\mu}\delta^{\nu ]}_{\lambda} \Big)+ \Theta^{\mu\nu}{}{}_{\lambda}=0
\end{gather}
where 
\begin{subequations}
\ba
 H^{\mu\nu}{}{}_{\lambda} & \equiv & a_{1}Q^{\nu\mu}{}{}_{\lambda}+2 a_{2}(Q^{\mu\nu}{}{}_{\lambda}+Q_{\lambda}{}^{\mu\nu})+2 b_{1}S^{\mu\nu}{}{}_{\lambda}+2 b_{2}S_{\lambda}{}^{[\mu\nu]}
 +c_{1}( S^{\nu\mu}{}{}_{\lambda}-S_{\lambda}{}^{\nu\mu}+Q^{[\mu\nu]}{}{}_{\;\lambda} \\
k_{\mu} & \equiv & 4 a_{3}Q_{\mu}+2 a_{5}q_{\mu}+2 c_{2}S_{\mu} \\
h_{\mu} & \equiv & a_{5} Q_{\mu}+2 a_{4}q_{\mu}+c_{3}S_{\mu} \\
f_{\mu} & \equiv & c_{2} Q_{\mu}+ c_{3}q_{\mu}+2 b_{3}S_{\mu} 
\ea
\end{subequations}
and
\be
\Theta^{\mu\nu}{}{}_{\lambda}  \equiv  \frac{\partial}{\partial \Gamma^{\lambda}{}_{\mu\nu}}\Big( \lambda g^{\alpha\beta}D_{\alpha}\psi D_{\beta}\psi \Big)
\ee
which for the conformally invariant case takes the form
\begin{subequations}
\beq
\Theta^{\mu\nu}{}{}_{\lambda} =-\lambda \left( \frac{n-2}{n} \right)\psi (D^{\nu}\psi)\delta^{\mu}_{\lambda}
\eeq
and for the frame rescaling invariant theory
\beq
\Theta^{\mu\nu}{}{}_{\lambda} =-2 \lambda \left( \frac{n-2}{n-1} \right)\psi (D^{[\mu}\psi)\delta^{\nu]}_{\lambda}
\eeq
\end{subequations}
since the gauge covariant derivative given by ($\ref{gcovQ}$) for the former and ($\ref{gcovS}$) for the latter respectively. To conclude, if we neglect the coupling via the covariant derivative, we obtain
\begin{subequations}
\begin{gather}
\psi^{2} \Big( H^{\mu\nu}{}{}_{\lambda}+\delta^{\mu}_{\lambda}k^{\nu}+\delta^{\nu}_{\lambda}h^{\mu}+g^{\mu\nu}h_{\lambda}+f^{[\mu}\delta^{\nu ]}_{\lambda} \Big)= 0 \label{gammaeqn3}
\end{gather}
however, for the conformally invariant case the $\Gamma$-field equations read
\begin{gather}
\psi^{2} \Big( H^{\mu\nu}{}{}_{\lambda}+\delta^{\mu}_{\lambda}k^{\nu}+\delta^{\nu}_{\lambda}h^{\mu}+g^{\mu\nu}h_{\lambda}+f^{[\mu}\delta^{\nu ]}_{\lambda} \Big)=\lambda \left( \frac{n-2}{n} \right)\psi (D^{\nu}\psi)\delta^{\mu}_{\lambda} \label{gammaeqn}
\end{gather}
and for the frame rescaling invariant case
\begin{gather}
\psi^{2} \Big( H^{\mu\nu}{}{}_{\lambda}+\delta^{\mu}_{\lambda}k^{\nu}+\delta^{\nu}_{\lambda}h^{\mu}+g^{\mu\nu}h_{\lambda}+f^{[\mu}\delta^{\nu ]}_{\lambda} \Big)=2 \lambda \left( \frac{n-2}{n-1} \right)\psi (D^{[\mu}\psi)\delta^{\nu]}_{\lambda}\,. \label{gammaeqn2}
\end{gather}
\end{subequations}
Now, to close the system of the field equations it remains to vary with respect to the scalar $\psi$. Neglecting the coupling via the covariant derivative, we would obtain simply
\begin{subequations}
\beq
\psi \mathcal{L}=\lambda\Box \phi \label{psi_hat}
\eeq
but in the conformally invariant case we find
\beq
\psi \mathcal{L}=\lambda \left( \frac{n-2}{4 n}Q_{\mu}D^{\mu}\psi+\frac{\partial_{\mu}(\sqrt{-g} D^{\mu}\psi)}{\sqrt{-g}}   \right) \label{psi_bar}
\eeq
while for the frame rescaling invariant theory, one obtains
\beq
\psi \mathcal{L}=\lambda \left( \frac{n-2}{n-1}S_{\mu}D^{\mu}\psi+\frac{\partial_{\mu}(\sqrt{-g} D^{\mu}\psi)}{\sqrt{-g}}   \right)\,. \label{psi_tilde}
\eeq
\end{subequations}
We have summarised the results in the Table \ref{theorytable} below.

\begin{table}[h]
\begin{tabular}{ |c||c|c|c|c|c||c|} 
 \hline
invariance & constraints  & $D_\mu = \partial_\mu - A_\mu$ & $g$-EoM & $\Gamma$-EoM & $\psi$-EoM & total \\
 \hline
 \hline
 projective &  Eq.(\ref{proinv}) & $A_{\mu}=0$ & Eq.(\ref{g_eom}) where $K_{\mu\nu}=0$ & Eq.(\ref{gammaeqn3})  &  Eq.(\ref{psi_hat}) & Eq.(\ref{res_hat})\\
\hline
 conformal & Eq.(\ref{confinv}) & $A_{\mu}=\left(\frac{n-2}{4n}\right)Q_{\mu} $ & Eq.(\ref{g_eom}) with (\ref{k_bar}) & Eq.(\ref{gammaeqn})  & Eq.(\ref{psi_bar}) & Eq.(\ref{res_bar})\\
 \hline
 frame rescaling & Eq.(\ref{frinv}) & $A_{\mu}=\left(\frac{n-2}{n-1}\right)S_{\mu}$ & Eq.(\ref{g_eom}) where $K_{\mu\nu}=0$  & Eq.(\ref{gammaeqn2})  & Eq.(\ref{psi_tilde}) & Eq.(\ref{res_tilde})\\
\hline
\end{tabular}
 \caption{The invariant actions (given by (\ref{plus}) by taking into account the constraints and the prescription for the derivative) and the equations of motion. The final column refers to the parameter
 constraints on the total action (\ref{total}) to be derived in Section \ref{general}. \label{theorytable}}
\end{table}

Let us examine (\ref{gammaeqn}) and (\ref{gammaeqn2}) a little further. To do so, notice that we can consider three operations on (\ref{gammaeqn}) and (\ref{gammaeqn2}). We can contract in $\mu=\lambda$, contact in $\nu=\lambda$ and multiply (and contact) by $g^{\mu\nu}$. Then we get three vector equations that we may formally write as
\begin{subequations}
\begin{gather}
\alpha_{1}Q_{\mu}+\alpha_{2}q_{\mu} +\alpha_{3}S_{\mu}=\frac{\partial_{\mu}\psi}{\psi}  \\
\beta_{1}Q_{\mu}+\beta_{2}q_{\mu} +\beta_{3}S_{\mu}=\frac{\partial_{\mu}\psi}{\psi}\\
\gamma_{1}Q_{\mu}+\gamma_{2}q_{\mu} +\gamma_{3}S_{\mu}=\frac{\partial_{\mu}\psi}{\psi}
\end{gather}
\end{subequations}
where the $\alpha_{i},\beta_{i},\gamma_{i}$ are all combinations of $a_{i},b_{i},c_{i}$ and $\lambda$. Then the above system of equations can be formally solved\footnote{Assuming that the determinant of the matrix corresponding to the system does not vanish.} to give
\beq
Q_{\mu}=\lambda_{1} \frac{\partial_{\mu}\psi}{\psi}\, q_{\mu}=\lambda_{2} \frac{\partial_{\mu}\psi}{\psi}\,S_{\mu}=\lambda_{3} \frac{\partial_{\mu}\psi}{\psi}
\eeq
where  the $\lambda_{i}'s$ depend on $\alpha_{i},\beta_{i},\gamma_{i}$. This result, when substituted back at (\ref{gammaeqn}) and (\ref{gammaeqn2}), yields
\beq
H^{\mu\nu}{}{}_{\lambda}=\sigma_{1}\delta^{\mu}_{\lambda}\frac{\partial_{\nu}\psi}{\psi}+\sigma_{2}\delta^{\nu}_{\lambda}\frac{\partial_{\mu}\psi}{\psi}+\sigma_{3}g^{\mu\nu}\frac{\partial^{\lambda}\psi}{\psi}
\eeq
where again the $\sigma_{i}'s$ depend on $a_{i},b_{i},c_{i}$ and $\lambda$. These manipulations make it clear that just as in the simple case studied in Section \ref{simple}, the non-metricity and torsion remain pure gauge in the case of the generic quadratic action which does not include the derivatives of the connection.

\section{The General Quadratic Theory}
\label{general}

In this Section we shall first complete the analysis of the previous Section by incorporating the parity-odd invariants. Then, in subsection \ref{appl}, we comment on possible applications
of our new results,
in particular we point out the various different frameworks wherein it could be interesting to build scale-invariant theories: considering post-Riemannian corrections to (a scale-free version of) the Einstein-Hilbert action, giving kinetic terms to the connection, or constructing scale-invariant teleparallel theories. 

\subsection{The Parity-Odd Scalars}

We now specialise to $n=4$ and take into account also the possible CP-violating terms, which were derived in the Appendix \ref{scalars}. We may thus write the action as 
\be
S = \frac{1}{2 \kappa}\int \diff^4 x \sqrt{-g}\Big[ \psi^{2}\lp \mathcal{L}_{Q} +  \mathcal{L}_{QT} + \mathcal{L}_{T}\rp +\lambda g^{\mu\nu}D_{\mu}\psi D_{\nu}\psi + \psi^4 \Lambda\Big]\,, \label{total}
\ee
where, for generality one could consider the quartic self-interaction (that in the $\psi=1$ gauge is seen just as the cosmological constant, and can be neglected in the following as irrelevant to our discussion),
and the three pieces of Lagrangians are now understood as $\mathcal{L}_{Q} = \mathcal{L}_{Q}^+ + \mathcal{L}_{Q}^-$,  $\mathcal{L}_{T} = \mathcal{L}_{T}^+ + \mathcal{L}_{T}^-$,
and $\mathcal{L}_{QT} = \mathcal{L}_{QT}^+ + \mathcal{L}_{QT}^-$, where the parity-even contributions had been specified in (\ref{even_s}), and the parity-odd contributions are given as follows:
\begin{subequations}
\ba
\mathcal{L}_{Q}^{-} & \equiv & a_{6}\epsilon^{\alpha\beta\gamma\delta}Q_{\alpha\beta\mu}Q_{\gamma\delta}{}^{\mu}=a_{6}A_{6} \\
\mathcal{L}_{T}^{-} & \equiv & b_{5}S_{\mu}t^{\mu}+b_{6}\epsilon^{\alpha\beta\gamma\delta}S_{\alpha\beta\mu}S_{\gamma\delta}{}^{\mu} =b_{5}B_{5}+b_{6}B_{6} \\
\mathcal{L}_{QT}^{-} & \equiv & c_{4}Q_{\mu}t^{\mu}+c_{5}q^{\mu}t_{\mu}+c_{6}\epsilon^{\alpha\beta\gamma\delta}Q_{\alpha\beta\mu}S_{\gamma\delta}{}^{mu}=c_{4}C_{4}+c_{5}C_{5}+c_{6}C_{6}
\ea
\end{subequations}
Let us now find the parameter space for the above action to be invariant under each of the three transformations, starting with the conformal transformation (\ref{bar}). The newly added parity-odd terms transform as
\begin{gather}
\bar{A}_{6}=e^{-2\phi}A_{6}\,, \quad \bar{B}_{5}=e^{-2\phi}B_{5}\,, \quad \bar{B}_{6}=e^{-2\phi}B_{6}
\end{gather}
and 
\be
\bar{C}_{4}=e^{-2\phi}( C_{4}-8 t^{\mu}\partial_{\mu}\phi )\,, \quad
\bar{C}_{5}=e^{-2\phi}( C_{5}-2 t^{\mu}\partial_{\mu}\phi )\,, \quad
\bar{C}_{6}=e^{-2\phi}( C_{6}-2 t^{\mu}\partial_{\mu}\phi ) 
\ee
under a conformal metric transformation. Thus, the only mixed parity-odd terms transform non-covariantly. 
We may write simply that 
\begin{subequations}
\ba
\bar{\mathcal{L}}_{Q}^{-} & = & e^{-2\phi}\mathcal{L}_{Q}^{-} \\ 
\bar{\mathcal{L}}_{T}^{-} & = &e^{-2\phi}\mathcal{L}_{T}^{-} \\
\bar{\mathcal{L}}_{QT}^{-} & = & e^{-2\phi}\mathcal{L}_{Q}^{-}-2e^{-2\phi}t^{\mu}(\partial_{\mu}\phi )(4 c_{4}+c_{5}+c_{6})\,.
\ea
\end{subequations}
The transformation for the parity-even part of the Lagrangian we have already computed in the previous section. So, for the total action to be invariant under conformal transformations in $n=4$ we must have 
\begin{subequations}
\label{res_bar}
\ba
0 & = & 2a_{1}+ 8 a_{3}+ a_{5}  \\ 
0 & = & a_{2}+a_{4}+2 a_{5} \\ 
0 & = & 4 a_{1}+a_{2}+ 16 a_{3}+a_{4}+4 a_{5}  \\
0 & = & c_{1}+4 c_{2}+c_{3} \\
0 & = & 4c_{4}+c_{5}+c_{6}\,. 
\ea
\end{subequations}
Note that the first four constraints in the above are the ones we had derived previously for the pure parity-even Lagrangian and the last constraint is imposed on the parity-odd part. We should mention that the additional constraint establishes a relation only between the coefficients of the parity-odd terms and does not mix them with the parameters of the parity-even scalars. 

Now, under a frame rescaling (\ref{tilde}) the parity-odd parts transform as
\begin{subequations}
\ba
\tilde{\mathcal{L}}_{Q}^{-} & = & e^{-2\phi}\mathcal{L}_{Q}^{-}\\ 
\tilde{\mathcal{L}}_{T}^{-} & = & e^{-2\phi}\mathcal{L}_{T}^{-}-\frac{1}{2}e^{-2 \phi}t^{\mu}\partial_{\mu}\phi \Big( 3b_{5}+4 b_{6} \Big) \\
\tilde{\mathcal{L}}_{QT}^{-} & = & e^{-2\phi}\mathcal{L}_{Q}^{-}\,.
\ea
\end{subequations}
And for the total action to be invariant under the frame rescalings, the parameters must satisfy
\begin{subequations}
\label{res_tilde}
\ba
0 & = & 2 b_{1}-b_{2}+3b_{3}=0   \\
0 & = & c_{1}+3 c_{2} \\
0 & = & c_{1}-3 c_{3} \\
0 & = & 3b_{5} + 4 b_{6}\,.
\ea
\end{subequations}
Again, the first three constraints above are the same with the pure parity-even theory and the last one is imposed among the parameters of the parity-odd terms. 

Now, in order to study the parameter space for the projective invariant case, again there is no need for a scalar field $\psi$ to compensate for the invariance (and if it is included, it should not be considered charged
under the transformation).
By (\ref{hat}), the parity-odd parts transform according to
\begin{subequations}
\ba
\hat{\mathcal{L}}_{Q}^{-}& = &\mathcal{L}_{Q}^{-}\\
\hat{\mathcal{L}}_{T}^{-} & = & \mathcal{L}_{T}^{-}-\frac{3}{2}b_{5}t_{\mu}\xi^{\mu}-2b_{6}t_{\mu}\xi^{\mu} \\
\hat{\mathcal{L}}_{QT}^{-} & = & \mathcal{L}_{QT}^{-}+2t_{\mu}\xi^{\mu}( 4c_{4}+c_{5}+c_{6})
\ea
\end{subequations}
as can be easily checked. So, projective invariance of the total action is ensured if the parameters satisfy
\begin{subequations}
\label{res_hat}
\ba
0 & = & 4 (2 a_{1}+ 8 a_{3}+a_{5})-c_{1}-3c_{2} \\
0 & = & 4 (2 a_{2}+2  a_{4}+4 a_{5})+c_{1}-3c_{3}\\
0 & = & -2 b_{1}+b_{2} -3 b_{3}+2(c_{1} +4 c_{2}+c_{3}) \\
0 & = & 16 (4 a_{1}+ a_{2}+ 16 a_{3}+ a_{4}+ 4 a_{5}) 
-3\Big( 2 b_{1}-b_{2}-3 b_{3}-4 (c_{1}+4 c_{2}+c_{3})\Big) \\
0 & = & -3b_{5}-4b_{6}+16c_{4}+4c_{5}+4c_{6}\,.
\ea
\end{subequations}
Note that in comparison with the pure parity-even case, the first four constraints remain the same, and a fifth additional constraint is imposed only among the parameters of the parity-odd scalars. The important thing is that the constraints again do not mix the parameters of the parity-even with the parameters of the parity-odd scalars.

Having completed the derivation of the scale-invariant theories, let us have a brief look at their generic properties. Let $T_{\mu\nu}$ be the stress-energy tensor and $H^\alpha{}_{\mu\nu}$ the
hypermomentum tensor for the matter fields. The two sets of field equations in any Palatini theory can then be written as
\be
M_{\mu\nu} = T_{\mu\nu}
\ee
and 
\be
\Xi^\alpha{}_{\mu\nu} = H^\alpha{}_{\mu\nu}\,.
\ee
As it is shown in detail in the Appendix \ref{identities}, the three versions of scale symmetry imply certain properties for these tensors. The conformal symmetry is associated with tracelessness.
That is, for the theory to be conformally invariant, a necessary requirement is that $M \equiv M^\alpha{}_\alpha=T^\alpha{}_\alpha \equiv T =0$. A basic property of the Maxwell field and the massless fermion
is that their energy-momentum tensors are traceless. Of course, the Proca field and the massive fermion break scale invariance by introducing the mass scale. Now, it is interesting to note that the projective symmetry on the other hand implies the tracelessness of the hypermomentum, in particular that $\Xi^\mu \equiv \Xi_\alpha{}^{\alpha\mu}=H_\alpha{}^{\alpha\mu}\equiv H^\mu=0$. The fermion is projectively invariant, but the Maxwell (or Proca) field is invariant only under the symmetric teleparallel projection (when we assume the minimal coupling principle $\partial_\mu \rightarrow \nabla_\mu$ of the Palatini formalism). Finally, the invariance of the theory under the frame rescalings implies the identities
\be
M = -\frac{\partial_\mu (\sqrt{-g}\Xi^\mu)}{2\sqrt{-g}} \quad \Rightarrow \quad T =  -\frac{\partial_\mu (\sqrt{-g} H^\mu)}{2\sqrt{-g}}\,.
\ee  
Interestingly, this version of scale symmetry can be compatible with a matter source that has a trace, given that the matter source also possesses hypermomentum.

\subsection{On Applications to Theory}
\label{appl}

As discussed in the introduction Section \ref{intro}, there is a vast amount of possible applications for scale-invariant theories. In fact, some may contemplate whether this symmetry should be
finally promoted to the same foundational status as the Lorentz symmetry, or perhaps even more properly the general coordinate invariance, to which it is in some sense related as the multiplication to the addition. In any case, in this paper, we have focused on formal developments, and will return to specific applications elsewhere.

However, it is pertinent to clarify in some detail how our results might be useful in various contexts of gravitational theory model building. At face value, our quadratic actions have trivial predictions.
Since we have not added kinetic terms to the connections, the quadratic action ${\mathcal{L}}_{Q}+{\mathcal{L}}_{T}+ {\mathcal{L}}_{QT}$ for any choice of parameters is but a generalised mass term, something like $\sim \Gamma^2$. This does give the connection any dynamics, but the solutions for the $\Gamma$ are pure-gauge, as was the case for the distortion in the explicitly-solved
simple example of Section \ref{simple}. Indeed we had sketched how one arrives at the same conclusion in the end of Section \ref{quadratic}. Now, one can of course add the linear curvature action of Section \ref{simple} to the quadratic action, and then the connection does get dynamics, those of GR\footnote{If there is hypermomentum, the situation changes, though the connection remains undynamical, as indeed is known from the seminal example of the Einstein-Cartan-Sciama-Kibble theory. Adding non-minimal derivative interactions of the connection to the matter sector 
 could of course make the connection dynamical.}. The reason is that though the curvature includes derivatives of the connection, they become mere boundary terms in the case of the action
 that is linear in the curvature. However, non-trivial dynamics could be achieved by taking into account quadratic invariants of curvature (which, we recall, are all invariant in $n=4$ under the three types of scale gauge transformations). In the context of considering corrections to the scale-invariant version of the Einstein-Cartan theory, the ``Einstein-Cartan-Weyl-Dirac'' action (\ref{confthe}), it seems to be a perfectly natural to include quadratic curvature terms besides the quadratic contribution $\psi^2\lp{\mathcal{L}}_{Q}+{\mathcal{L}}_{T}+ {\mathcal{L}}_{QT}\rp$.
 
On the other hand, we are now fully armed with the scale-invariant arsenal to set the torsion and the non-metricity propagating directly by including their derivatives into the action. In the previous subsection we had completed the derivation of the covariant scalars, and the covariant derivatives acting on scalars we had already deduced in Section \ref{simple}. From these ingredients, we can construct non-integrable scale-invariant geometries, was that what we desired. In such a case, we would incorporate kinetic terms for the gauge field $A_\mu$ (which, recall, is the non-metricity trace a.k.a. Weyl vector in the case of conformal and the torsion trace in the case of frame rescaling invariance). 

However, it can be even more interesting to constrain the dynamics of the connection, instead of (or perhaps, in addition to) adding kinetic terms via curvature terms or covariant derivatives. In teleparallism, we supplement the action using the Lagrange multiplier $\lambda_\alpha{}^{\beta\mu\nu}$ that transforms covariantly under both the general coordinate and the scale transformation.
The term $\lambda_\alpha{}^{\beta\mu\nu}R^\alpha{}_{\beta\mu\nu}$ then does not break the invariance, but appropriately restricts the rotational part of connection to be pure gauge, i.e. forces the $\Gamma$ to be flat. In the metric-compatible
case, we should add a further Lagrange multiplier that sets the non-metricity tensor to vanish, and it turns out that, quite interestingly, even without adding the explicit derivatives, the torsion obtains its dynamics via such a mechanism\footnote{An alternative method is the Golovnev's ``inertial variation'' \cite{Golovnev:2017dox} wherein instead of the connection one varies a gauge transformation parameter. However, it is against the spirit of the Palatini formalism to set the connection a priori into the purely inertial form.}
 \cite{BeltranJimenez:2017tkd,BeltranJimenez:2018vdo}. It was known that there is a unique parameter combination that yields the teleparallel equivalent of GR, and from the results of this paper we see that there is a 3-parameter class of models that is invariant under frame rescalings, and a full 5-parameter class of models that is conformally invariant 
(in both cases, we are not counting the overall normalisation of the action but taking into account the parameter $\lambda$). Note that the frame rescaling has to be understood now in its teleparallel version reported in Table \ref{table1} i.e. only the  antisymmetric part of the connection enjoys the projection.
 
On the other hand, one may augment the flatness constraint with the constraint of vanishing torsion, leading then to symmetric teleparallelism, see \cite{Conroy:2017yln,Jarv:2018bgs,Runkla:2018xrv,Hohmann:2018xnb,Hohmann:2018wxu,Soudi:2018dhv,Adak:2018vzk} for current studies into such geometry. This is a totally different perspective to the theory of gravity, wherein the spacetime affine 
connection can be actually fully eliminated, as was only recently clarified in \cite{BeltranJimenez:2017tkd,BeltranJimenez:2018vdo}. We may thus in principle ``purify'' gravitation from inertial
forces, and the possibly profound implications of this insight certainly call for further investigation. It was known that the parameter combination of the non-metricity scalars that leads to the equivalent of GR in such a geometry is unique, and that it is exclusively for this combination that the affine connection, to the linear order, in addition to being pure gauge, decouples from the action \cite{BeltranJimenez:2017tkd}.
From the results of this paper we may add that this combination, as in fact any other, is also covariant under the frame rescaling wherein only the symmetric part of the connection
undergoes a projection (this was referred to as projection$_\parallel$ in Table \ref{table1}). However, under the conformal transformation (\ref{bar}) there exists only a two-parameter class of covariant scalars which is second order in the derivatives of the metric. Promoting the covariant scalars into invariant ones with the aid of the dilaton and including the associated parameter $\lambda$, one can easily verify that the symmetric teleparallel equivalent of GR is included amongst the conformally invariant quadratic theories.

\section{Conclusion}
\label{conclu}

\begin{table}[t]
\begin{tabular}{ |c|c|c|c|c|} 
 \hline
 covariance  &  $ Q^2$  & $T^2$ & $QT$ & $\sum$   \\ \hline
 coordinate +  & $A_1$, $A_2$, $A_3$, $A_4$, $A_5$ & $B_1$, $B_2$, $B_3$ & $C_1$, $C_2$, $C_3$ & 11 \\ 
 coordinate -  & $A_6$ & $B_5$, $B_6$ & $C_4$, $C_5$, $C_6$ &  6  \\ \hline 
 projective +  &\multicolumn{3}{|c|}{$(A_1-\frac{1}{n}A_3)$, $(A_2-A_4)$, $(B_1+2B_2)$,
 $(B_1-\frac{2}{n-1}B_3)$, $(C_1-\frac{1}{2}C_2)$, $(\frac{1}{8n}A_1-\frac{1}{8}A_2+C_1-\frac{1}{n}C_2)$, $(\frac{n-1}{8}A_2-B_1-C_3)$} & 7 \\ 
 projective -  & $A_6$ &\multicolumn{2}{|c|}{ $(B_6+4C_4)$, $(B_6+C_5)$, $(B_6+C_6)$, $(4B_5+3C_6)$ } & 5 \\ \hline 
 conformal  + & $(nA_1-A_3)$, $(A_2-A_4)$ & $B_1$, $B_2$, $B_3$ & $(C_1-C_3)$, $(nC_1-C_2)$ & 7 \\ 
 conformal -  & $A_6$ & $B_5$, $B_6$ & $(C_4-4C_5)$, $(C_4-4C_6)$ & 5 \\ \hline
 rescaling + & $\qquad$ $A_1$, $A_2$, $A_3$, $A_4$, $A_5$ $\qquad$ & $\qquad$ $(B_1+2B_2)$, $(B_2+\frac{1}{n-1}B_3)$ $\qquad$ & $(n-1)C_1-C_2+C_3$  & 8  \\ 
 rescaling - & $A_6$  & $(4B_5-3B_6)$ & $C_4$, $C_5$, $C_6$ & 5  \\ 
 \hline
\end{tabular}
 \caption{The scalars and co-covariants in the three versions of rescalings, separated according to parity. (The linearly independent combinations are not unique of course). In the odd-parity cases, $n=4$. \label{table2}} 
\end{table}

After a century since its introduction, we considered it timely to revisit the formulation of a scale-invariant theory, in particular in view of the consistent and viable theories that may be 
constructed from invariants of non-metricity, torsion, and both.

Scale transformations in metric-affine geometry have been considered previously. For example, in Section 6.1 of the review  \cite{Hehl:1994ue}, in addition to the curvature
sector which is more trivial, the basic properties of the quadratic non-metric scalars and torsion scalars are clarified, i.e. that the torsion trace squared has a different covariance property than the tensorial and the axial invariant, and the volume-changing non-metricity invariant has a different covariance property than the volume-preserving invariants. The points of departure in this paper were that 0) we presented the analysis in the Palatini formalism. Though the exterior calculus is elegant and makes some aspects of the analysis more transparent, probably more workers in the field are fluent in the tensor formalism, which on the other hand is indeed more straightforward for some of the practical applications. We have also 1) further clarified the geometric interpretations of the relevant versions of scale transformations and presented the systematical analysis of all of them in a unified framework. Concrete generalisations of the results in the previous literature are that we have 2) taken into account the possible couplings between torsion and non-metricity and 3) included the CP-violating terms. The map to the main results was given in the Table \ref{theorytable}, and furthermore, in Table \ref{table2}, we present a summary of the invariants we have derived in Sections \ref{quadratic} and \ref{general}.

\appendix

\section{The Quadratic Non-Metricity and Torsion Scalars}
\label{scalars}

In this Appendix we derive the quadratic invariants of torsion and non-metricity. First we write systematically down all the possible contraction of the tensors (\ref{q}) and (\ref{torsion}), and then sort out the 
independent ones. We end up with the same set of scalars as in e.g. Ref.\cite{Pagani:2015ema}.

Let us thus first list the relevant scalars. The pure non-metricity scalars are
\begin{subequations}
\begin{gather}
A_{1}=Q_{\alpha\mu\nu}Q^{\alpha\mu\nu} \\
A_{2}=Q_{\alpha\mu\nu}Q^{\mu\nu\alpha} \\
A_{3}=Q_{\mu}Q^{\mu} \\
A_{4}=q_{\mu}q^{\mu} \\
A_{5}=Q_{\mu}q^{\mu} \\
A_{6}=\epsilon^{\alpha\beta\gamma\delta}Q_{\alpha\beta\mu}Q_{\gamma\delta}{}^{\mu}
\end{gather}
\end{subequations}
where the two independent traces are defined\footnote{We are using $q_{\mu}=Q_{\alpha\mu\nu}g^{\alpha\nu}$ for the second non-metricity vector, instead of $\tilde{Q}_\mu$ \cite{BeltranJimenez:2018vdo,Jarv:2018bgs} to avoid confusion that may appear from the various symbols.} as $Q_{\alpha}\equiv Q_{\alpha\mu\nu}g^{\mu\nu}$ and $q_{\mu}=Q_{\lambda\nu\mu}g^{\lambda\nu}$.
Note that there appear to be five independent quadratic even invariants, though only four irreducible components of the the non-metricity tensor (the binom, conom, vecnom and conom, see the Appendix B1 of Ref. \cite{Hehl:1994ue}). 

The pure torsion scalars one may write down are
\begin{subequations}
\begin{gather}
B_{1}=S_{\alpha\mu\nu}S^{\alpha\mu\nu} \\
B_{2}=S_{\alpha\mu\nu}S^{\mu\nu\alpha} \\
B_{3}=S_{\mu}S^{\mu}
\\
B_{4}=t_{\mu}t^{\mu}
\\
B_{5}=S_{\mu}t^{\mu}
\\
B_{6}=\epsilon^{\alpha\beta\gamma\delta}S_{\alpha\beta\mu}S_{\gamma\delta}{}^{\mu}
\\
B_{7}=\epsilon^{\alpha\beta\gamma\delta}S_{\lambda\alpha\beta}S^{\lambda}{}_{\gamma\delta}
\\
B_{8}=\epsilon^{\alpha\beta\gamma\delta}S_{\mu\alpha\beta}S_{\gamma\delta}{}^{\mu}
\end{gather}
\end{subequations}
where $S_{\mu}\equiv S_{\mu\lambda}{}^{\lambda}$ and $t^{\alpha}\equiv \epsilon^{\alpha\beta\gamma\delta}S_{\beta\gamma\delta}$.
It should be noted that $B_4$ is not independent of the $B_1$, $B_2$ and $B_3$, and only two of the four pseudoscalars $B_5$, $B_6$, $B_7$ and $B_8$ is independent. Let first us show the redundancy of $B_{4}=t_{\mu}t^{\mu}$. By a direct calculation, this is found to be
\ba
B_{4} & = & t_{\mu}t^{\mu}=\epsilon_{\mu\alpha\beta\gamma}\epsilon^{\mu\kappa\lambda\rho}S^{\alpha\beta\gamma}S_{\kappa\lambda\rho}=-3! \delta^{[\kappa}_{\alpha}\delta^{\lambda}_{\beta}\delta^{\rho]}_{\gamma}S^{\alpha\beta\gamma}S_{\kappa\lambda\rho} =
-3! S^{\alpha\beta\gamma}S_{[\alpha\beta\gamma]} \nonumber \\ & = & -2 S^{\alpha\beta\gamma}\lp S_{\alpha\beta\gamma}+S_{\gamma\beta\alpha}+S_{\beta\gamma\alpha}\rp   
=-2(B_{1}+2B_{2})\,.
\ea
Thus we may discard $B_4$ in the following without loss of generality. Let us then consider the parity-odd torsion scalars in $n=4$. We start from the definition of $t^\rho$,
which when  contracted by $\epsilon_{\rho\alpha\beta\mu}$ and using $\epsilon^{\rho\kappa\lambda\sigma}\epsilon_{\rho\alpha\beta\mu}=-3! \delta^{[\kappa}_{\alpha}\delta^{\lambda}_{\beta}\delta^{\sigma]}_{\mu}$ gives
\beq
\epsilon_{\rho\alpha\beta\mu}t^{\rho}=-3! S_{[\alpha\beta\mu]}\,.
\eeq
Exploiting the antisymmetry of the torsion tensor in its first two indices the above may be expressed as
\beq
\epsilon_{\rho\alpha\beta\mu}t^{\rho}=-2(  S_{\alpha\beta\mu}+S_{\mu\alpha\beta} +S_{\beta\mu\alpha}    )\,.
\eeq
Furthermore, contracting the above with $\epsilon^{\alpha\beta\gamma\delta}$ and using $\epsilon_{\rho\alpha\beta\mu}\epsilon^{\alpha\beta\gamma\delta}=-4 \delta^{[\gamma}_{\rho}\delta^{\delta]}_{\mu}$ we finally arrive at
\beq
2 t^{[\gamma}\delta^{\delta]}_{\mu}=\epsilon^{\alpha\beta\gamma\delta}S_{\alpha\beta\mu}+2 \epsilon^{\alpha\beta\gamma\delta} S_{\mu\alpha\beta}\,. \label{tepseq}
\eeq
The latter is the key equation that gives the relations among the parity-odd terms. To obtain these, we first contract ($\ref{tepseq}$) by $S_{\gamma\delta}{}^{\mu}$ and use the definitions of $B_{i}'s$ to obtain
\beq
2 B_{5}=B_{6}+2B_{8}
\eeq
In addition, contracting with $S^{\mu}{}_{\gamma\delta}$ this time, gives
\beq
-B_{5}=B_{8}+2 B_{7}
\eeq
Therefore, we have two equations relating the $B_{5},...,B_{8}$ and so only two of the four are independent. We may choose the $B_{5}$ and $B_{6}$.

Finally, it is possible to form scalars by mixing non-metricity and torsion. Such invariants are
\begin{subequations}
\begin{gather}
C_{1}=Q_{\alpha\mu\nu}S^{\alpha\mu\nu}
\\
C_{2}=Q_{\mu}S^{\mu} 
\\
C_{3}=q_{\mu}S^{\mu}
\\
C_{4}=Q^{\mu}t_{\mu}
\\
C_{5}=q^{\mu}t_{\mu}
\\
C_{6}=\epsilon^{\alpha\beta\gamma\delta}Q_{\alpha\beta\mu}S_{\gamma\delta}{}^{\mu}
\\
C_{7}=\epsilon^{\alpha\beta\gamma\delta}Q_{\alpha\beta\mu}S^{\mu}{}_{\gamma\delta}
\end{gather}
\end{subequations}
Again, there is redundancy in parity-odd terms. Out of the four combinations $C_{4}, C_{5},C_{6},C_{7}$ only the three are independent. This is easily seen by contracting (\ref{tepseq}) with $Q_{\gamma\delta}{}^{\mu}$ to arrive at
\beq
C_{4}-C_{5}=C_{6}+2 C_{7}\,.
\eeq
Therefore one scalar is redundant and we choose to disregard $C_{7}$.  

This exhausts the list of the quadratic second order scalars in metric-affine geometry. Further reason, besides that their transformation properties are trivial, that we need not consider the curvature invariants beyond the 
$R=g^{\mu\nu}R^{\alpha}{}_{\mu\alpha\nu}$ and the $\epsilon^{\alpha\beta\gamma\delta}R_{\alpha\beta\gamma\delta}$, is that by
by decomposing the connection into the metric (\ref{christoffel}), the non-metric part (known often as disformation) and the torsion part (known often as contortion) we can always rewrite all the curvature terms of the metric invariants and the above scalars (and their higher derivatives, which we leave out of from the present analysis). Note also that the linear scalars $\nabla_\mu T^\mu$, $\nabla_\mu Q^\mu$ and $\nabla_\mu q^\mu$ are
redundant, up to boundary terms. 

A remark about the parity-odd terms is in order. While our construction is the most general for any dimensions for the parity-even terms, it is stricly speaking restricted to $n=4$ when considering the parity-odd terms. The reason is that in $n$ dimensions one would have at hand the $n$-dimensional totally antisymmetric Levi-Civita symbol, which is a technical complication. Here we restrict to using the symbol only with four indices. 

\subsection{On quartic invariants}

Having established the transformation laws for the quadratic torsion and non-metricity scalars in Section \ref{quadratic}, we may us now find some (of the many!) quartic combinations that remain invariant, for example, under conformal metric transformations. To start with, let us first note that
\begin{gather}
(n \bar{A}_{1}-\bar{A}_{3})=e^{-2 \phi}(n A_{1}-A_{3}) \nonumber \\
( \bar{A}_{2}-\bar{A}_{4})=e^{-2 \phi}( A_{2}-A_{4}) \nonumber \\
\Big( \bar{A}_{5}-\frac{n}{2}\bar{A}_{4} -\frac{1}{2n}\bar{A}_{3}\Big)=e^{-2 \phi}\Big( A_{5}-\frac{n}{2}A_{4} -\frac{1}{2n} A_{3}\Big) \nonumber \\
\bar{B}_{i}=e^{-2 \phi}B_{i}\, \quad \forall \,\, i \nonumber \\
( \bar{C}_{1}-\bar{C}_{3})=e^{-2 \phi}(C_{1}-C_{3}) \nonumber \\
(n \bar{C}_{1}-\bar{C}_{2})=e^{-2 \phi}( n C_{1}-C_{2}) \nonumber \\
(n \bar{C}_{3}-\bar{C}_{2})=e^{-2 \phi}( n C_{3}-C_{2}) \nonumber \\
(2 \bar{C}_{2}-n \bar{C}_{1}-n \bar{C}_{3})=e^{-2 \phi}( 2 C_{2}- n C_{1}-n C_{3}) \nonumber 
\end{gather} 
under $\bar{g}_{\mu\nu}=e^{2\phi}g_{\mu\nu}$. This in turn means that any of the above combinations when squared or multiplied by another combination of the list, yields a conformally invariant scalar. For instance,
in $n=4$,
\beq
\sqrt{-g}(4 A_{1}-A_{3})^{2}
\eeq
\beq
\sqrt{-g}( A_{2}-A_{4})B_{2}
\eeq
are both conformally invariant. In total we can form $7^2=49$ conformal invariants from the squares of the even-parity quadratic covariant combinations. However, the total number of conformal invariants is probably larger, since there are more scalars one can form by contracting the indices 4 tensors than the square of the number of scalars formed by contracting the indices of two tensors.

\section{Variational identities}
\label{identities}

In this Section, we will derive some preliminary results which will be helpful in the rest of this paper. First, we derive some necessary variational formulae. As our aim to to construct scale-invariant theories,
we will elucidate the generic relations between three versions of scale-invariance and the tracelessness properties of the variational terms (without yet specifying the particular actions).  

\subsection{Variations}

Let us gather here the various variations that we will use in what follows. We start with torsion and compute variations with respect to the metric first. We have
\beq
T_{\mu\nu\lambda}(\delta_{g}S^{\mu\nu\lambda})=\delta g^{\mu\nu}\Big( T_{\mu\alpha\beta}S_{\nu}{}^{\alpha\beta}-T_{\alpha\nu\beta}S_{\mu}{}^{\alpha\beta}\Big)=\delta g^{\mu\nu}(2T_{[\nu\alpha]\beta}S_{\mu}{}^{\alpha\beta})
\eeq
and also
\beq
T^{\mu\nu\lambda}(\delta_{g}S_{\mu\nu\lambda})=-\delta g^{\mu\nu} \Big( T^{\alpha\beta}{}{}_{\nu}S_{\alpha\beta\mu} \Big)
\eeq
where $T_{\mu\nu\lambda}$ is an arbitrary tensor field (or tensor density). Then setting $T_{\mu\nu\lambda}=S_{\mu\nu\lambda}$ one has
\beq
S_{\mu\nu\lambda}(\delta_{g}S^{\mu\nu\lambda})=\delta g^{\mu\nu}(2S_{\nu\alpha\beta}S_{\mu}{}^{\alpha\beta})
\eeq
as well as
\beq
S^{\mu\nu\lambda}(\delta_{g}S_{\mu\nu\lambda})=-\delta g^{\mu\nu}(S_{\alpha\beta\mu}S^{\alpha\beta}{}{}_{\nu})
\eeq
such that
\beq
\delta_{g}(S_{\mu\nu\lambda}S^{\mu\nu\lambda})=\delta g^{\mu\nu}\Big( 2S_{\nu\alpha\beta}S_{\mu}{}^{\alpha\beta}-S_{\alpha\beta\mu}S^{\alpha\beta}{}{}_{\nu} \Big)\,.
\eeq
Therefore, setting $T_{\mu\nu\lambda}=S_{\lambda\mu\nu}$ we can conclude that
\beq
\delta_{g}(S_{\mu\nu\lambda}S^{\lambda\mu\nu})=-S_{\nu\alpha\beta}S_{\mu}{}^{\alpha\beta}(\delta g^{\mu\nu})\,.
\eeq
Now, using
\beq
\delta_{g}\epsilon_{\alpha\beta\gamma\delta}=\delta_{g}(\sqrt{-g}\eta_{\alpha\beta\gamma\delta})=-\frac{1}{2}\epsilon_{\alpha\beta\gamma\delta} g_{\mu\nu}\delta  g^{\mu\nu}
\eeq
we compute
\beq
A^{\alpha}\delta_{g}t_{\alpha}=\delta g^{\mu\nu} \left[ -\frac{1}{2}g_{\mu\nu}A_{\alpha}t^{\alpha}+2 A^{\lambda}\epsilon_{\lambda\nu\alpha\beta}S_{\mu}{}^{\alpha\beta} \right]
\eeq
where $A^{\mu}$ is an arbitrary vector. Then, also using that $\delta_{g}S_{\mu}=0$ we find
\beq
\delta_{g}(t_{\alpha}S^{\alpha})=\delta g^{\mu\nu} \left[ -\frac{1}{2}g_{\mu\nu}S_{\alpha}t^{\alpha}+2 S^{\lambda}\epsilon_{\lambda\nu\alpha\beta}S_{\mu}{}^{\alpha\beta} +t_{\mu}S_{\nu}\right]
\eeq
and also 
\beq
\delta_{g}(S_{\alpha}S^{\alpha})=\delta g^{\mu\nu}( S_{\mu}S_{\nu})\,.
\eeq
Following the same procedure for the rest of the quadratic torsion scalars, we finally derive the metric variations
\begin{gather}
\delta_{g}B_{1}=\delta g^{\mu\nu}\Big( 2S_{\nu\alpha\beta}S_{\mu}{}^{\alpha\beta}-S_{\alpha\beta\mu}S^{\alpha\beta}{}{}_{\nu} \Big) \nonumber \\
\delta_{g}B_{2}=\delta g^{\mu\nu}(-S_{\nu\alpha\beta}S_{\mu}{}^{\alpha\beta}) \nonumber \\
\delta_{g}B_{3}=\delta g^{\mu\nu}( S_{\mu}S_{\nu}) \nonumber \\
\delta_{g}B_{5}=\delta g^{\mu\nu} \left[ -\frac{1}{2}g_{\mu\nu}S_{\alpha}t^{\alpha}+2 S^{\lambda}\epsilon_{\lambda\nu\alpha\beta}S_{\mu}{}^{\alpha\beta} +t_{\mu}S_{\nu}\right] \nonumber \\
\delta_{g}B_{6}=\delta g^{\mu\nu}\left( \frac{1}{2}g_{\mu\nu}B_{6}-\epsilon^{\alpha\beta\gamma\delta}S_{\alpha\beta\mu}S_{\gamma\delta\nu} \right)\nonumber \\
\delta_{g}B_{7}=\delta g^{\mu\nu}\left(\frac{1}{2}g_{\mu\nu}B_{7}+2 S^{\alpha\beta}{}{}_{\mu}S_{\alpha}{}^{\gamma\delta}\epsilon_{\nu\beta\gamma\delta}+\epsilon^{\alpha\beta\gamma\delta}S_{\mu\alpha\beta}S_{\nu\gamma\delta} \right) \nonumber \\
\delta_{g}B_{8}=\delta g^{\mu\nu}\left( \frac{1}{2}g_{\mu\nu}B_{8}-\epsilon^{\beta}{}_{\nu\gamma\delta}S^{\gamma\delta\alpha}S_{\mu\alpha\beta} \right)\,. \nonumber
\end{gather}
Finally, we shall also need the variations of with respect to the connection. For the $\Gamma$-variations of non-metricity scalars we find
\begin{gather}
\delta_{\Gamma}A_{1}=\delta_{\Gamma}(Q_{\alpha\mu\nu}Q^{\alpha\mu\nu})=(4 Q^{\nu\mu}{}{}_{\lambda} ) \delta \Gamma^{\lambda}{}_{\mu\nu} \nonumber \\
\delta_{\Gamma}A_{2}=\delta_{\Gamma}(Q_{\alpha\mu\nu}Q^{\mu\nu\alpha})=2 (Q^{\mu\nu}{}{}_{\lambda} +Q_{\lambda}{}^{\mu\nu})\delta \Gamma^{\lambda}{}_{\mu\nu} \nonumber \\
\delta_{\Gamma}A_{3}=\delta_{\Gamma}(Q_{\mu}Q^{\mu})=(4 Q^{\nu} \delta^{\mu}{}{}_{\lambda} ) \delta \Gamma^{\lambda}{}_{\mu\nu} \nonumber \\
\delta_{\Gamma}A_{4}=\delta_{\Gamma}(\tilde{Q}_{\mu} \tilde{Q}^{\mu})=2(\tilde{Q}_{\lambda}g^{\mu\nu}+\tilde{Q}^{\mu}\delta_{\lambda}^{\nu}) \delta \Gamma^{\lambda}{}_{\mu\nu} \nonumber \\
\delta_{\Gamma}A_{5}=\delta_{\Gamma}( Q_{\mu} \tilde{Q}^{\mu})=( 2 \tilde{Q}^{\nu}\delta_{\lambda}^{\mu}+ Q_{\lambda}g^{\mu\nu}+Q^{\mu}\delta^{\nu}_{\lambda}) \delta \Gamma^{\lambda}{}_{\mu\nu}\,. \nonumber 
\end{gather}
The $\Gamma$-variations of the torsion are straightforward. We are now armed with the formulas that allow to readily obtain the field equations for an arbitrary metric-affine theory consisting of the scalars in Section \ref{scalars}. 

\subsection{Projective Invariance and Tracelessness}

As we have already pointed out in the previous Section \ref{simple}, the Palatini Tensor
\begin{equation}
P_{\lambda}{}^{\mu\nu}\equiv \frac{\delta R}{\delta \Gamma^{\lambda}{}_{\mu\nu}}
\end{equation}
has zero trace when contracted in its first two indices, that is
\begin{equation}
P_{\mu}{}^{\mu\nu}=0\,.
\end{equation}
In fact as we have argued before, any tensor constructed out of a projective invariant quantity has this property. Let us prove this here. Consider the scalar quantity
$\Psi$ that is invariant under projective transformations. Then define
\begin{equation}
\Xi_{\lambda}{}^{\mu\nu}\equiv \frac{\delta \Psi}{\delta \Gamma^{\lambda}{}_{\mu\nu}}\,.
\end{equation}
Now consider the projective transformation $\Gamma^{\lambda}{}_{\mu\nu}\longrightarrow \hat{\Gamma}^{\lambda}{}_{\mu\nu} =\Gamma^{\lambda}{}_{\mu\nu}+ \delta_{\mu}^{\lambda}\xi_{\nu}$ 
such that\footnote{{$\hat{\delta}$} denotes a projective variation of the connection, and in the following $\bar{\delta}$ will denote the variation under the conformal transformation and $\tilde{\delta}$ the variation under
the frame rescaling.}
\begin{equation}
\hat{\delta}\Gamma^{\lambda}{}_{\mu\nu}=\hat{\Gamma}^{\lambda}{}_{\mu\nu} -\Gamma^{\lambda}{}_{\mu\nu}= \delta_{\mu}^{\lambda}\xi_{\nu}
\end{equation}
Applying the latter transformation to $\psi$, we have
\begin{equation}
\hat{\delta}\psi =\frac{\delta \psi}{\hat{\delta}\Gamma^{\lambda}{}_{\mu\nu}} \hat{\delta}\Gamma^{\lambda}{}_{\mu\nu}=\Xi_{\lambda}{}^{\mu\nu}\delta_{\mu}^{\lambda}\xi_{\nu}=\Xi_{\mu}{}^{\mu\nu}\xi_{\nu}
\end{equation}
Now since $\psi$ is invariant, we have that $\hat{\delta}\psi=0$. Thus, using this, along with the fact that the vector $\xi_{\nu}$ is arbitrary, from the above we conclude that 
\begin{equation}
\Xi_{\mu}{}^{\mu\nu}=0
\end{equation}
as we had stated.

\subsection{Conformal Invariance and Tracelessness}

As we have proven above, if a scalar quantity is invariant under projective transformations then its variation with respect to the connection yields a tensor (or tensor density if we do not divide the result by $\sqrt{-g}$) that is traceless in its first two indices. Similarly, if a scalar density (which we may integrate to construct an action of course) is invariant under conformal transformations then its variation with respect to the metric tensor yields a tensor that is traceless. Let us prove this here.\newline
\textbf{Proof:} Consider the scalar density $\sqrt{-g} \Psi \label{psig}$,  
where $\Psi$ is again a scalar. Then define the variation
\beq
M_{\mu\nu} \equiv \frac{1}{\sqrt{-g}}\frac{\delta (\sqrt{-g} \Psi)}{\delta g^{\mu\nu}}
\eeq
and denote its trace by $M\equiv M_{\mu\nu}g^{\mu\nu}$.
Consider now a conformal transformation of the metric $\bar{g}_{\mu\nu}=e^{2 \phi}g_{\mu\nu}$, or in its contravariant form
$\bar{g}^{\mu\nu}=e^{-2 \phi}g^{\mu\nu}$. 
Expanding the latter for infinitesimal transformations, it follows that
\beq
\bar{g}^{\mu\nu} \approx (1-2\phi)  g^{\mu\nu} \Rightarrow \bar{\delta}g^{\mu\nu}=-2 \phi g^{\mu\nu}
\eeq
where $\bar{\delta}g^{\mu\nu}\equiv \bar{g}^{\mu\nu}-g^{\mu\nu} $ denotes the infinitesimal change the metric undergoes under the conformal transformation. Given that $\sqrt{-g} \Psi$ is invariant under conformal transformations we have 
\begin{gather}
 \bar{\delta}(\sqrt{-g} \Psi)=0\Rightarrow M_{\mu\nu}\bar{\delta}g^{\mu\nu}=0\Rightarrow \nonumber
 -2\phi M_{\mu\nu}g^{\mu\nu}=0
\end{gather}
and since the last one must hold true for arbitrary $\phi$ we conclude that
\beq
M=M_{\mu\nu}g^{\mu\nu}=0
\eeq
as stated.\newline
\textbf{Examples:} Let us confirm the strength of the above statement with two examples. First consider the scalar density in $4$ dimensions\footnote{This of course generalizes to any dimension and takes the form $\sqrt{-g}R^{\frac{n}{2}}$ where $n$ is the dimension of the space.}
\beq
\sqrt{-g}R^{2}
\eeq
which is conformally invariant as can be easily seen. Its metric variation is found to be
\beq
M_{\mu\nu}=\frac{1}{\sqrt{-g}}\frac{\delta (\sqrt{-g}R^{2})
}{\delta g^{\mu\nu}}=2 R\left( R_{\mu\nu}-\frac{1}{4}g_{\mu\nu}R\right)
\eeq
and therefore
\beq
M=M_{\mu\nu}g^{\mu\nu}=2R\left( R-R\right)=0
\eeq
as expected. As a second example consider
\beq
\sqrt{-g}R_{\mu\nu}R^{\mu\nu}
\eeq
which is also a conformally invariant quantity in $4$ dimensions (as we recall, is any possible curvature-squared term). Variation with respect to the metric yields 
\begin{gather}
M_{\mu\nu}=\frac{1}{\sqrt{-g}}\frac{\delta (\sqrt{-g}R_{\mu\nu}R^{\mu\nu})
}{\delta g^{\mu\nu}} 
=-\frac{1}{2}R_{\alpha\beta}R^{\alpha\beta}g_{\mu\nu}+R_{\mu}{}^{\beta}R_{\nu\beta}+R^{\beta}{}_{\nu}R_{\beta\mu}
\end{gather}
which again gives a vanishing trace since
\beq
M=M_{\mu\nu}g^{\mu\nu}=-2R_{\mu\nu}R^{\mu\nu}+R_{\mu\nu}R^{\mu\nu}+R_{\mu\nu}R^{\mu\nu}=0\,.
\eeq
Note that from these considerations apply also for any matter source. It is indeed very well-known that the conformal symmetry is compatible only with traceless matter sources, the Maxwell field being the 
seminal example. 

\subsection{The case of Frame Rescalings}

As we have seen a frame rescaling results in a conformal transformation $+$ a special projective transformation both powered by a single scalar field $\phi(x)$.\footnote{This is most important because one can also have projective and conformal transformations that are powered by different fields. Then invariance means that both metric and connection conjugates have zero traces and they are not related. As an example consider $\sqrt{-g}R^{2}$ which is independently invariant under $ \Gamma^{\lambda}{}_{\mu\nu} \rightarrow\Gamma^{\lambda}{}_{\mu\nu}+ \delta_{\mu}^{\lambda}\xi_{\nu} $  and $g_{\mu\nu} \rightarrow e^{2 \phi}g_{\mu\nu}$ (where $\xi_{\nu}$ and $\phi$ are not related to any way) and as a result $M=0$ and $\Xi^{\mu}=0$.} We will now prove that if a scalar density is invariant under frame rescalings then the trace of its metric conjugate and the divergence of the trace of its connection conjugate are related to one another.\newline
\textbf{Proof:} Consider the action
\beq
S=\int \diff^{n}x \sqrt{-g}\Psi \label{Theor}
\eeq
and recall the definitions of the metric and connection conjugates
\beq
M_{\mu\nu} \equiv \frac{1}{\sqrt{-g}}\frac{\delta (\sqrt{-g} \Psi)}{\delta g^{\mu\nu}}\,, \quad 
\Xi_{\lambda}{}^{\mu\nu}\equiv \frac{1}{\sqrt{-g}}\frac{\delta (\sqrt{-g} \mathcal{L})}{\delta \Gamma^{\lambda}{}_{\mu\nu}}=\frac{\delta \mathcal{L}}{\delta \Gamma^{\lambda}{}_{\mu\nu}}
\end{equation}
and the definitions of the traces
\beq
M=g^{\mu\nu}M_{\mu\nu}\,, \quad \Xi^{\mu} =\Xi_{\lambda}{}^{\lambda\mu}\,.
\eeq
We now state that if ($\ref{Theor}$) is invariant under frame rescalings then
\beq
2M+\frac{\partial_{\mu}(\sqrt{-g}\Xi^{\mu})}{\sqrt{-g}}=0\,.
\eeq
To prove this let us compute the change in ($\ref{Theor}$) under frame rescalings. Using $\tilde{\delta}g^{\mu\nu}=\tilde{g}^{\mu\nu}-g^{\mu\nu}=-2\phi g^{\mu\nu}$ and
$\tilde{\delta}\Gamma^{\lambda}{}{}_{\mu\nu}=\tilde{\Gamma}^{\lambda}{}{}_{\mu\nu}-\Gamma^{\lambda}{}{}_{\mu\nu}=\delta_{\mu}^{\lambda}\partial_{\nu}\phi$
where $\tilde{\delta}$ denotes the change under frame rescalings, we compute
\ba
\tilde{\delta}S & = & \int \diff^{n}x \Big[ \sqrt{-g}M_{\mu\nu}\tilde{\delta}g^{\mu\nu}+\sqrt{-g}\Xi_{\lambda}{}^{\mu\nu}\tilde{\delta}\Gamma^{\lambda}{}{}_{\mu\nu}\Big]   
=\int \diff^{n}x \Big[ \sqrt{-g}(-2\phi g^{\mu\nu}M_{\mu\nu})+\sqrt{-g}\Xi_{\lambda}{}^{\mu\nu}\delta_{\mu}^{\lambda}\partial_{\nu}\phi \Big] \nonumber \\
& = & \int \diff^{n}x \Big[ -\sqrt{-g}2 \phi M +\sqrt{-g}\Xi^{\mu}\partial_{\mu}\phi \Big] 
 =-\int \diff^{n}x \Big[ \sqrt{-g}2 \phi M +\phi(\partial_{\mu}\sqrt{-g}\Xi^{\mu}) \Big]+\int \diff^{n}x \partial_{\mu}( \sqrt{-g}\Xi^{\mu}\phi) \nonumber \\
& = & -\int \diff^{n}x  \phi \Big[ \sqrt{-g}2 M +(\partial_{\mu}\sqrt{-g}\Xi^{\mu}) \Big] + \text{a boundary term}\,.
\ea
Then, ignoring surface terms, since $S$ is invariant it follows that 
\beq
\tilde{\delta}S\Rightarrow \phi \Big[ \sqrt{-g}2 M +(\partial_{\mu}\sqrt{-g}\Xi^{\mu}) \Big]=0
\eeq
and since the last equality must be true for arbitrary $\phi$ we conclude that
\beq
2M+\frac{\partial_{\mu}(\sqrt{-g}\Xi^{\mu})}{\sqrt{-g}}=0
\eeq
as stated. Notice that if the above result is applied for the matter part of the action, for frame rescaling invariant matter the traces of the energy momentum and hyper-momentum tensors are related through
\beq
2T+\frac{\partial_{\mu}(\sqrt{-g}\Delta^{\mu})}{\sqrt{-g}}=0
\eeq
where $T=T_{\mu\nu}g^{\mu\nu}$ and $\Delta^{\nu}=\Delta_{\mu}{}^{\mu\nu}=0$

\section{Transformation identities}

In this Section we will investigate the transformation properties of the torsion tensor, nonmetricity tensor and the related quantities under the three versions of rescalings.

\subsection{Projective Transformations}
\label{b_hat}

Recall that the projective transformation is $\Gamma^{\lambda}{}_{\mu\nu}\rightarrow \hat{\Gamma}^{\lambda}{}_{\mu\nu}=\Gamma^{\lambda}{}_{\mu\nu}+\delta^{\lambda}_{\mu}\xi_{\nu}$,
$g_{\mu\nu}\rightarrow \hat{g}_{\mu\nu}=g_{\mu\nu}$. 
It is easy to see that under the above, the torsion and non-metricity tensors transform as
\beq
\hat{S}_{\mu\nu}{}^{\lambda}=S_{\mu\nu}{}^{\lambda}+\delta^{\lambda}_{[\mu}\xi_{\nu]} \quad 
\text{or} \quad
 \hat{S}_{\mu\nu\alpha}=S_{\mu\nu\alpha} +g_{\alpha[\mu}\xi_{\nu]} \label{tortr}
\eeq
and
\beq
\hat{Q}_{\alpha\mu\nu}= Q_{\alpha\mu\nu}+2 \xi_{\alpha}g_{\mu\nu}\,.
\eeq
For their associate vectors we have
\begin{equation}
\hat{S}_{\mu} = S_{\mu} + \frac{(1-n)}{2}\xi_{\mu}\,, \quad 
\hat{Q}_{\mu}= Q_{\mu} +2 n \xi_{\mu}\,, \quad 
\text{and}
\hat{q}_{\mu}=q_{\mu} +2 \xi_{\mu}\,.
\end{equation}
Note that this implies
\beq
S_{[\mu\nu\alpha]}\longrightarrow S_{[\mu\nu\alpha]} 
\eeq
the projective invariance of the totally antisymmetric torsion. It is this part of torsion that couples to fermions. Thus, the Dirac action is projectively invariant, in addition to being invariant under the frame rescalings.

\subsection{Conformal Transformations}
\label{b_bar}

What we call the conformal transformation in this paper is defined by $g_{\mu\nu}\rightarrow \bar{g}_{\mu\nu}=e^{2 \phi}g_{\mu\nu}$ and $\Gamma^{\lambda}{}_{\mu\nu}\rightarrow \bar{\Gamma}^{\lambda}{}_{\mu\nu}=\Gamma^{\lambda}{}_{\mu\nu}$. That is, under a conformal transformation the metric tensor picks up a conformal factor $e^{2 \phi} $ while the affine connection is left unchanged. Note that the contravariant form of the metric tensor transforms as $\bar{g}^{\mu\nu}=e^{-2\phi}g^{\mu\nu}$ as can be easily seen from the relation $g_{\mu\nu}g^{\nu\lambda}=\delta_{\mu}^{\lambda}$. In addition, the square root of the determinant of the metric obeys the transformation rule
\beq
\sqrt{-\bar{g}}=e^{n \phi} \sqrt{-g}
\eeq 
and for $n=4$ thus $\sqrt{-\bar{g}}=e^{4 \phi} \sqrt{-g}$, which is obtained directly by first taking the determinant of $\bar{g}_{\mu\nu}$ and then taking the square root of the result. From this last relation we infer the transformation rule for the Levi-Civita tensor
\beq
\bar{\epsilon}_{\mu\nu\rho\sigma}=e^{4 \phi}\epsilon_{\mu\nu\rho\sigma}
\eeq
\beq
\bar{\epsilon}^{\mu\nu\rho\sigma}=e^{-4 \phi}\epsilon^{\mu\nu\rho\sigma}
\eeq
and recall that $\epsilon_{\mu\nu\rho\sigma}=\sqrt{-g}\eta_{\mu\nu\rho\sigma}$ where $\eta_{\mu\nu\rho\sigma}$ is the Levi-Civita symbol. Using the above we see that torsion and non-metricity transform as
\beq
\bar{S}_{\mu\nu}{}^{\lambda}=S_{\mu\nu}{}^{\lambda}
\eeq
\beq
\bar{Q}_{\alpha\mu\nu}=e^{2 \phi}\Big( Q_{\alpha\mu\nu}-2(\partial_{\alpha} \phi)g_{\mu\nu} \Big)
\eeq
and the related vectors as
\begin{gather}
\bar{Q}_{\mu}=Q_{\mu}-2 n \partial_{\mu} \phi\,,
\quad
\bar{q}_{\mu}=q_{\mu}-2 \partial_{\mu}\phi
\end{gather}
and
\begin{gather}
\bar{S}_{\mu}=S_{\mu}\,,
\quad
\bar{t}_{\mu}=t_{\mu}\,.
\end{gather}
Then, it follows that all pure torsion scalars
\begin{subequations}
\begin{gather}
B_{1}=S_{\alpha\mu\nu}S^{\alpha\mu\nu}\,, \,\,
B_{2}=S_{\alpha\mu\nu}S^{\mu\nu\alpha}\,, \,\,
B_{3}=S_{\mu}S^{\mu}\,, \,\,
B_{4}=t_{\mu}t^{\mu} \label{even_t}
\end{gather}
\end{subequations}
and
\begin{subequations}
\begin{gather}
B_{5}=S_{\mu}t^{\mu}\,, \,\,
B_{6}=\epsilon^{\alpha\beta\gamma\delta}S_{\alpha\beta\mu}S_{\gamma\delta}{}^{\mu}\,, \,\,
B_{7}=\epsilon^{\alpha\beta\gamma\delta}S_{\lambda\alpha\beta}S^{\lambda}{}_{\gamma\delta}\,, \,\,
B_{8}=\epsilon^{\alpha\beta\gamma\delta}S_{\mu\alpha\beta}S_{\gamma\delta}{}^{\mu} \label{odd_t}
\end{gather}
\end{subequations}
are all conformally covariant, that is
\beq
\bar{B}_{i}=e^{-2 \phi}B_{i}
\eeq
for any $i=1,2,...,8$. This means that any combination $\sqrt{-g}B_{i}B_{j}$ is conformal invariant when $n=4$.
Regarding the pure non-metricity scalars, one can verify the transformation laws
\begin{subequations}
\begin{gather}
\bar{A}_{1}=\bar{Q}_{\alpha\mu\nu}\bar{Q}^{\alpha\mu\nu}=e^{-2 \phi}\Big[ A_{1}-4 Q^{\mu}\partial_{\mu} \phi +4 n (\partial \phi)^{2}\Big]\\
\bar{A}_{2}=\bar{Q}_{\alpha\mu\nu}\bar{Q}^{\mu\nu\alpha}=e^{-2 \phi}\Big[ A_{2}-4 q^{\mu}\partial_{\mu} \phi  +4  (\partial \phi)^{2}\Big]
 \\
\bar{A}_{3}=\bar{Q}_{\mu}\bar{Q}^{\mu}=e^{- 2\phi}\Big[ A_{3}-4 n Q^{\mu}\partial_{\mu}\phi +4 n^{2} (\partial \phi)^{2}\Big] 
\\
\bar{A}_{4}=\bar{q}_{\mu}\bar{q}^{\mu}=e^{- 2\phi}\Big[ A_{4}-4  q^{\mu}\partial_{\mu}\phi +4  (\partial \phi)^{2}\Big] 
\\
\bar{A}_{5}=\bar{Q}_{\mu}\bar{q}^{\mu}=e^{- 2\phi}\Big[ A_{5}-2(Q^{\mu} +n  q^{\mu})\partial_{\mu}\phi +4 n (\partial \phi)^{2}\Big] 
\\
\bar{A}_{6}=\bar{\epsilon}^{\alpha\beta\gamma\delta}\bar{Q}_{\alpha\beta\mu}\bar{Q}_{\gamma\delta}{}^{\mu}=e^{-2 \phi}A_{6}
\end{gather}
\end{subequations}
and for the mixed terms 
\begin{subequations}
\begin{gather}
\bar{C}_{1}=\bar{Q}_{\alpha\mu\nu}\bar{S}^{\alpha\mu\nu}=e^{-2 \phi}\Big[C_{1} - 2  S^{\mu}\partial_{\mu}\phi \Big]
\\
\bar{C}_{2}=\bar{Q}_{\mu}\bar{S}^{\mu} =e^{-2 \phi}\Big[C_{2} - 2 n S^{\mu}\partial_{\mu}\phi \Big]
\\
\bar{C}_{3}=\bar{q}_{\mu}\bar{S}^{\mu}=e^{-2 \phi}\Big[C_{3} - 2  S^{\mu}\partial_{\mu}\phi \Big]
\\
\bar{C}_{4}=\bar{Q}^{\mu}\bar{t}_{\mu}=e^{-2 \phi}\Big[C_{4} - 2 n  t^{\mu}\partial_{\mu}\phi \Big]
\\
\bar{C}_{5}=\bar{q}^{\mu}\bar{t}_{\mu}=e^{-2 \phi}\Big[C_{5} - 2  t^{\mu}\partial_{\mu}\phi \Big]
\\
\tilde{C}_{6}=\bar{\epsilon}^{\alpha\beta\gamma\delta}\bar{Q}_{\alpha\beta\mu}\bar{S}_{\gamma\delta}{}^{\mu}=e^{-2 \phi}\Big[C_{6} - 2  t^{\mu}\partial_{\mu}\phi \Big]
\\
\bar{C}_{7}=\bar{\epsilon}^{\alpha\beta\gamma\delta}\bar{Q}_{\alpha\beta\mu}\bar{S}^{\mu}{}_{\gamma\delta}=e^{-2 \phi}\Big[C_{7} - 2  t^{\mu}\partial_{\mu}\phi \Big]\,.
\end{gather}
\end{subequations}

\subsection{Frame Rescaling}
\label{b_tilde}

A frame rescaling transformation results in a combination of  a conformal metric transformation$+$ a special projective transformation. More specifically, we have
$\Gamma^{\lambda}{}_{\mu\nu}\rightarrow \tilde{\Gamma}^{\lambda}{}_{\mu\nu}=\Gamma^{\lambda}{}_{\mu\nu}+\delta^{\lambda}_{\mu}\partial_{\nu}\phi$
and $g_{\mu\nu}\rightarrow \tilde{g}_{\mu\nu}=e^{2 \phi}g_{\mu\nu}$,
with the same scalar field $\phi(x)$ appearing in both above. Interestingly, under the this transformation, the non-metricity tensor does not change and it just picks-up a conformal factor.  In words
\beq
\tilde{Q}_{\alpha\mu\nu}=e^{2\phi}Q_{\alpha\mu\nu}
\eeq
as can be easily seen by applying both transformations on $(\ref{q})$. This makes the procedure  of computing quadratic non-metricity scalars, extremely simple. Indeed, let us recall the scalars
\begin{subequations}
\begin{gather}
A_{1}=Q_{\alpha\mu\nu}Q^{\alpha\mu\nu}\,, \,\,
A_{2}=Q_{\alpha\mu\nu}Q^{\mu\nu\alpha}\,, \,\,
A_{3}=Q_{\mu}Q^{\mu}\,, \,\,
A_{4}=q_{\mu}q^{\mu}\,, \,\,
A_{5}=Q_{\mu}q^{\mu}\,, \,\,
A_{6}=\epsilon^{\alpha\beta\gamma\delta}Q_{\alpha\beta\mu}Q_{\gamma\delta}{}^{\mu}\,.
\end{gather}
\end{subequations}
It is an easy matter to show that under frame rescalings
\beq
\tilde{A}_{i}=e^{-2\phi}A_{i}
\eeq
for any $i=1,2,...,6$. Therefore any combination $\sqrt{-g}A_{i}A_{j}$ is invariant under frame rescalings.  As far as torsion is concerned, we have the transformation law
\beq
\tilde{S}_{\mu\nu}{}^{\lambda}=S_{\mu\nu}{}^{\lambda}+\delta^{\lambda}_{[\mu}\partial_{\nu]}\phi
\eeq
and for the torsion vector
\beq
\tilde{S}_{\mu}=S_{\mu}+\frac{(1-n)}{2}\partial_{\mu}\phi\,.
\eeq
Then, considering the even-parity torsion scalars (\ref{even_t})
we see that under a frame rescaling, they transform as 
\begin{subequations}
\begin{align}
\tilde{B}_{1}=e^{-2\phi} \Big[ B_{1}-2S^{\mu}\partial_{\mu}\phi
+\frac{(n-1)}{2} (\partial \phi)^{2}\Big] \\
\tilde{B}_{2}=e^{-2\phi}\Big[ B_{2}+S^{\mu}\partial_{\mu}\phi +\frac{(1-n)}{4}(\partial \phi)^{2} \Big] \\
\tilde{B}_{3}=e^{-2\phi}\Big[B_{3}+(1-n)S^{\mu} \partial_{\mu}\phi +\frac{(1-n)^{2}}{4}(\partial \phi)^{2} \Big]
\end{align}
\end{subequations}
Notice that the combinations $B_{1}+2B_{2}$, $\, (n-1)B_{1}-2B_{3}$ and $(n-1)B_{2}+B_{3}$ transform covariantly. For the odd-parity quadratic torsion scalars (\ref{odd_t}) one finds
\begin{subequations}
\begin{gather}
\tilde{B}_{4}=e^{-2\phi} B_{4}
\\
\tilde{B}_{5}=e^{-2\phi} \Big[ B_{5}+\frac{(1-n)}{2}t^{\mu}\partial_{\mu}\phi \Big]
\\
\tilde{B}_{6}=e^{-2\phi} \Big[ B_{6}-2 t^{\mu}\partial_{\mu}\phi \Big]
\\
\tilde{B}_{7}=e^{-2\phi} \Big[ B_{7}+ t^{\mu}\partial_{\mu}\phi \Big]
\\
\tilde{B}_{8}=e^{-2\phi} \Big[ B_{8}-\frac{1}{2} t^{\mu}\partial_{\mu}\phi \Big]
\end{gather}
\end{subequations}

For the mixed terms one finds
\begin{subequations}
\begin{gather}
\tilde{C}_{1}=\tilde{Q}_{\alpha\mu\nu}\tilde{S}^{\alpha\mu\nu}=e^{-2\phi} \Big[ C_{1}+\frac{1}{2}(q^{\mu}-Q^{\mu})\partial_{\mu}\phi \Big]
\\
\tilde{C}_{2}=\tilde{Q}_{\mu}\tilde{S}^{\mu}=e^{-2\phi} \Big[ C_{2}+\frac{1}{2}(1-n)Q^{\mu}\partial_{\mu}\phi \Big]
\\
\tilde{C}_{3}=\tilde{q}_{\mu}\tilde{S}^{\mu}=e^{-2\phi} \Big[ C_{3}+\frac{1}{2}(1-n)q^{\mu}\partial_{\mu}\phi \Big]
\\
\tilde{C}_{4}=\tilde{Q}_{\mu}\tilde{t}^{\mu}=e^{-2\phi}C_{4}
\\
\tilde{C}_{5}=\tilde{q}_{\mu}\tilde{t}^{\mu}=e^{-2\phi}C_{5}
\\
\tilde{C}_{6}=\tilde{\epsilon}^{\alpha\beta\gamma\delta}\tilde{Q}_{\alpha\beta\mu}\tilde{S}_{\gamma\delta}{}^{\mu}=e^{-2 \phi}C_{6}
\\
\tilde{C}_{7}=\tilde{\epsilon}^{\alpha\beta\gamma\delta}\tilde{Q}_{\alpha\beta\mu}\tilde{S}^{\mu}{}_{\gamma\delta}=e^{-2 \phi}C_{6}
\end{gather}
\end{subequations}

\textbf{Example:} As an example consider the scalar density
\beq
\sqrt{-g}A^{2}=\sqrt{-g}(Q_{\mu}Q^{\mu})^{2}
\eeq
which is invariant under frame rescalings in $n=4$  as can be easily seen. Its metric conjugate reads (where we have dropped a total derivative that is assumed to vanish on the boundary)
\beq
M_{\mu\nu}=-\frac{1}{2}g_{\mu\nu}A^{2}+2A Q_{\mu}Q_{\nu}-4 g_{\mu\nu}\frac{\partial_{\alpha}{(\sqrt{-g}Q^{\alpha}A)}}{\sqrt{-g}}
\eeq
with the trace
\beq
M=-16\frac{\partial_{\alpha}{(\sqrt{-g}Q^{\alpha}A)}}{\sqrt{-g}}
\eeq
The associated connection conjugate is found to be
\beq
\Xi_{\lambda}{}^{\mu\nu}=8AQ^{\nu}\delta^{\mu}_{\lambda}
\eeq
with the trace (in the first two indices)
\beq
\Xi^{\nu}=32AQ^{\nu}\,. 
\eeq
So we observe that
\beq
\frac{\partial_{\mu}(\sqrt{-g}\Xi^{\mu})}{\sqrt{-g}}=32\frac{\partial_{\alpha}{(\sqrt{-g}Q^{\alpha}A)}}{\sqrt{-g}}=-2M \quad \Rightarrow \quad
2M+\frac{\partial_{\mu}(\sqrt{-g}\Xi^{\mu})}{\sqrt{-g}}=0
\eeq
as expected.

\bibliography{transformations}

\end{document}